\documentclass[letterpaper,english,aps,prb,amsmath,amssymb,reprint,author-year,author-numerical,floatfix,raggedfooter,superscriptaddress]{revtex4-1}
\usepackage[latin9]{inputenc}
\usepackage{mathtools}
\usepackage{bm}
\usepackage{amsmath}
\usepackage{amssymb}

\usepackage{graphicx}

\usepackage[caption=false]{subfig}
\usepackage{verbatim}

\makeatletter

\providecommand{\tabularnewline}{\\}

\usepackage{dcolumn}
\usepackage{bm}
\expandafter\let\csname equation*\endcsname\relax
\expandafter\let\csname endequation*\endcsname\relax
\usepackage{booktabs}

\usepackage{xcolor}
\usepackage{soul}

\makeatother

\usepackage{babel}
\begin{document}

\preprint{AIP/123-QED}

\title{Theoretical studies of electronic transport in mono- and bi-layer phosphorene:\\
 A critical overview}

\author{Gautam Gaddemane}

\affiliation{Department of Materials Science and Engineering, The University of
Texas at Dallas~\\
 800 W. Campbell Rd., Richardson, TX 75080, USA}

\author{William G. Vandenberghe}

\affiliation{Department of Materials Science and Engineering, The University of
Texas at Dallas~\\
 800 W. Campbell Rd., Richardson, TX 75080, USA}
 
 \author{Maarten L. Van de Put}

\affiliation{Department of Materials Science and Engineering, The University of
Texas at Dallas~\\
 800 W. Campbell Rd., Richardson, TX 75080, USA}
 
 \author{Shanmeng Chen}

\affiliation{Department of Materials Science and Engineering, The University of
Texas at Dallas~\\
 800 W. Campbell Rd., Richardson, TX 75080, USA}
 
 \author{Sabyasachi Tiwari}

\affiliation{Department of Materials Science and Engineering, The University of
Texas at Dallas~\\
 800 W. Campbell Rd., Richardson, TX 75080, USA}

\author{Edward Chen}

\affiliation{Corporate Research and Chief Technology Office, Taiwan Semiconductor Manufacturing Company Ltd.~\\
 168, Park Ave. II, Hsinchu Science Park, Hsinchu 300-75, Taiwan,
R.O.C.}

\author{Massimo V. Fischetti}

\affiliation{Department of Materials Science and Engineering, The University of
Texas at Dallas~\\
 800 W. Campbell Rd., Richardson, TX 75080, USA}
\email{max.fischetti@utdallas.edu.}

\selectlanguage{english}%

\date{\today}
\begin{abstract}
Recent \textit{ab initio} theoretical calculations of the electrical
performance of several two-dimensional materials predict a low-field
carrier mobility that spans several orders of magnitude (from 26,000
to 35~cm$^{2}$ V$^{-1}$ s$^{-1}$, for example, for the hole mobility
in monolayer phosphorene) depending on the physical approximations
used. Given this state of uncertainty, we review critically the physical
models employed, considering phosphorene, a group V material, as a specific example. We
argue that the use of the most accurate models results in a calculated
performance that is at the disappointing lower-end of the predicted
range. We also employ first-principles methods to study high-field
transport characteristics in mono- and bi-layer phosphorene.
For thin multi-layer phosphorene we confirm the most disappointing results,
with a strongly anisotropic carrier mobility that does not
exceed $\sim$ 30 cm$^{2}$ V$^{-1}$ s$^{-1}$ at 300 K for electrons along the armchair direction.
\end{abstract}

\maketitle

\section{Introduction}
\label{sec:Intro} 

In the past couple of decades, the theoretical study of electronic
transport in semiconductors has been affected by two new driving factors.
First, the effort to scale transistors to the nanometer-size has stimulated
interest in materials and devices that are quite different from the
`conventional' materials employed by the micro-electronics industry.
References~\onlinecite{Wang_2014,Houssa_2016b,Molle_2017} constitute
excellent recent overviews of the state of the art. Unlike silicon,
germanium, or III-V compound semiconductors, for which decades of
study have resulted in a reliable database of their electronic properties
({\em e.g.,} band gap, effective mass, and carrier mobility), the
atomic, electronic, and transport properties of many of these new
materials are, at best, poorly known; at worst, even their existence
and stability are known only from theoretical predictions. The infancy
of the technology used to deal with these materials also casts doubts
on the usefulness of experimental results, because of the large deviations
from ideality that are expected from such an immature and fast-changing
technology.

The second driving cause is the timely and welcome progress recently
made in \textit{ab initio} (or `first principles') theoretical methods.
Whereas in the past their predictions have been limited to small systems
and had little or no connection to electronic transport, recent progress
made in physical understanding, numerical algorithms, and computing
hardware has broadened their range of applications, improved their
accuracy, and extended their scope to electronic transport.\cite{DFT_review}
Density functional theory (DFT) is now routinely used to predict the
atomic and electronic structure of these new materials, thanks to
the wide availability of computer packages, such as the Vienna \textit{Ab~initio}
Software Package (VASP)\cite{VASP1,VASP2,VASP3,VASP4} or Quantum Espresso (QE)\cite{Giannozzi_2009} .
Even the strength of the electron-phonon interaction can now be calculated
using DFT by using either finite ion displacements\cite{Vandenberghe_2015} or Density function Perturbation Theory (DFPT)\cite{Baroni_2001,Giustino_2017}, a remarkable evolution since the early `pioneering' days in which the rigid-ion approximation\cite{Ziman_1958}
and empirical pseudopotentials were painstakingly used to estimate
deformation potentials in Si, intervalley
deformation potentials in III-V compound semiconductors\cite{Zollner_1991,Zollner_1992},
and used in Monte Carlo transport studies\cite{Fischetti_1991,Yoder_1993}.
Even transport in open systems has been studied using DFT\cite{Choi_1999}
and such an \textit{ab initio} formalism has also been used to study
dissipative transport in the two-dimensional materials of current
interest\cite{Gunst_2016}. 

Despite this remarkable progress, and limiting ourselves to the carrier
mobility in covalent two-dimensional materials, theoretical predictions
reported in the literature disagree wildly. Our purpose here to analyze
critically the situation we face regarding phosphorene, taken as a
striking example of this uncertainty and disagreement that is due
to both physical and computational aspects, understand the underlying
causes, learn from this how we should proceed, and consider in detail
low-field and high-field electronic transport in phosphorene. 
Therefore, this paper is organized as follows:
In Sec.~\ref{sec:Case} we discuss the state-of-the-art regarding
the carrier mobility in phosphorene, presenting results obtained in
a simplified but realistic model. In Sec.~\ref{sec:Methods} we present
a general theoretical framework to study low- and high-field electronic
transport in 2D crystals using DFT. Finally, we present our results
for mono- and bi-layer phosphorene in Sec.~\ref{sec:phosphorene}.

\section{Carrier mobility in phosphorene}
\label{sec:Case} 

Monolayer black phosphorus (bP), or phosphorene, is one of the many
two-dimensional materials that have attracted enormous interest since
the isolation of graphene\cite{Geim_2007}. Considering only covalent
crystals, notable examples that we shall mention or consider
explicitly here include silicene\cite{Houssa_2011,Vogt_2012,Roome_2014,Tao_2015},
germanene\cite{Li_2013,Davila_2014}, phosphorene itself, of course\cite{Gomez_2014,Xia_2014,Li_2014,Liu_2014,Cao_2015,Doganov_2015,Xiang_2015,Gillgren_2015,Tayari_2015},
arsenene\cite{Kamal_2015,Li_2016,Pizzi_2016,Xu_2017}, antimonene\cite{Pizzi_2016,Xu_2017,Singh_2016,Ji_2016},
stanene\cite{Xu_2013,Zhu_2015,Suarez_2015}, and another large-band-gap
two-dimensional topological insulator, bismuthene\cite{Ford_1992,Whittle_1995} which, known since
the 1990s, has become the
subject of recent renewed interest\cite{Khomitsky_2014,Reis_2016}.
Interest in phosphorene presumably originates from the very large
carrier mobility measured in bulk black phosphorous\cite{Akahama_1983,Morita_1986}.
This interest has been reinforced by the good measured electrical
properties of field-effect transistors (FETs) having many-layer phosphorene
as channel material\cite{Xia_2014,Li_2014,Liu_2014,Cao_2015,Doganov_2015,Xiang_2015,Gillgren_2015,Tayari_2015}.
Despite such wide interest, to our knowledge, the intrinsic charge-transport
characteristics of monolayer phosphorene have not been widely studied
experimentally, having been reported only in Ref.~\onlinecite{Cao_2015}.
Moreover, theoretical predictions are in wild disagreement. We shall
now review the experimental and theoretical information available
at present, before discussing the causes of the theoretical confusion.
\begin{table}
\caption{Experimentally measured hole mobility, $\mu_{{\rm h}}$, in phosphorene
multilayers at 300~K.}
\begin{ruledtabular}
\begin{tabular}{ccccc}
Reference  & $\mu_{{\rm h}}$ (cm$^{2}$V$^{-1}$s$^{-1}$)  & thickness  &  & \tabularnewline
\hline 
Akahama \textit{et al.}$^{{\rm {(a)}}}$  & 150-1,300  & bulk black phosphorus  &  & \tabularnewline
Li \textit{et al.}$^{{\rm {(b)}}}$  & 300-1,000  & $>$ 10~nm  &  & \tabularnewline
Xia \textit{et al.}$^{{\rm {(c)}}}$  & 600  & 15~nm  &  & \tabularnewline
Gillgren \textit{et al.}$^{{\rm {(d)}}}$  & 400  & `few layers'  &  & \tabularnewline
Doganov \textit{et al.}$^{{\rm {(e)}}}$  & 189  & 10~nm  &  & \tabularnewline
Xia \textit{et al.}$^{{\rm {(c)}}}$  & 400  & 8~nm  &  & \tabularnewline
Liu \textit{et al.}$^{{\rm {(f)}}}$  & 286  & 5~nm  &  & \tabularnewline
Xiang \textit{et al.}$^{{\rm {(g)}}}$  & 214  & 4.8~nm  &  & \tabularnewline
Cao \textit{et al.}$^{{\rm {(h)}}}$  & 1  & monolayers  &  & \tabularnewline
Cao \textit{et al.}$^{{\rm {(h)}}}$  & 80  & bilayers  &  & \tabularnewline
Cao \textit{et al.}$^{{\rm {(h)}}}$  & 1,200  & trilayers  &  & \tabularnewline
\end{tabular}
\end{ruledtabular}

\begin{raggedright}
(a) Ref.~\onlinecite{Akahama_1983} \hspace*{1.5cm} (e) Ref.~\onlinecite{Doganov_2015}\\
 (b) Ref.~\onlinecite{Li_2014} \hspace*{1.5cm} (f) Ref.~\onlinecite{Liu_2014}\\
 (c) Ref.~\onlinecite{Xia_2014} \hspace*{1.5cm} (g) Ref.~\onlinecite{Xiang_2015}\\
 (d) Ref.~\onlinecite{Gillgren_2015} \hspace*{1.5cm} (h) Ref.~\onlinecite{Cao_2015} 
\par\end{raggedright}
\label{tab:mu_exp} 
\end{table}

\begin{table}
\caption{Theoretical calculations of the 300~K electron and hole mobility,
$\mu_{{\rm e}}$ and $\mu_{{\rm h}}$, in monolayer and bilayer phosphorene.}
\begin{ruledtabular}
\begin{tabular}{ccccccccc}
Reference  & \multicolumn{2}{c}{$\mu_{{\rm e}}$ (cm$^{2}$V$^{-1}$s$^{-1}$) } & \multicolumn{2}{c}{$\mu_{{\rm h}}$ (cm$^{2}$V$^{-1}$s$^{-1}$) } &  &  &  & \tabularnewline
 & armchair  & zigzag  & armchair  & zigzag  &  &  &  & \tabularnewline
\hline 
\multicolumn{5}{c}{monolayers} &  &  &  & \tabularnewline
\hline 
Qiao \textit{et al.}$^{{\rm {(a)}}}$  & 1,100  & 80  & 640-700  & 10,000-26,000  &  &  &  & \tabularnewline
Jin \textit{et al.}$^{{\rm {(b)}}}$  & 210  & 40  & 460  & 90  &  &  &  & \tabularnewline
Rudenko \textit{et al.}$^{{\rm {(c)}}}$  & 738  & 114  & 292  & 157  &  &  &  & \tabularnewline
Rudenko \textit{et al.}$^{{\rm {(d)}}}$  & $\sim$700  &  & $\sim$250  &  &  &  &  & \tabularnewline
Trushkov \textit{et al.}$^{{\rm {(e)}}}$  & 625  & 82  &  &  &  &  &  & \tabularnewline
Liao \textit{et al.}$^{{\rm {(f)}}}$  & 170  & 50  & 170  & 35  &  &  &  & \tabularnewline
This work$^{{\rm {(g)}}}$  & 20  & 10  & 19  & 2.4  &  &  &  & \tabularnewline
This work$^{{\rm {(h)}}}$  & 21  & 10  & 19  &  3 &  &  &  & \tabularnewline
This work$^{{\rm {(i)}}}$  & 25  & 5  &  &  &  &  &  & \tabularnewline
\hline 
\multicolumn{5}{c}{bilayers} &  &  &  & \tabularnewline
\hline 
Qiao \textit{et al.}$^{{\rm {(a)}}}$  & 600  & 140-160  & 2,600-2,800  & 1,300-2,200  &  &  &  & \tabularnewline
Jin \textit{et al.}$^{{\rm {(b)}}}$  & 1,020  & 360  & 1,610  & 760  &  &  &  & \tabularnewline
This work$^{{\rm {(g)}}}$  & 14  & 7  & 12  & 2  &  &  &  & \tabularnewline

\end{tabular}
\end{ruledtabular}

\begin{raggedright}
(a) Ref.~\onlinecite{Qiao_2014}\\
 (b) Ref.~\onlinecite{Jin_2016}, Monte Carlo and DFT (DFPT) \\
 (c) Ref.~\onlinecite{Rudenko_2016}, LA and TA, one-phonon processes\\
 (d) Ref.~\onlinecite{Rudenko_2016}, LA and TA, one- and two-phonon
processes\\
 (e) Ref.~\onlinecite{Trushkov_2017}, LA and TA, at a density of
$10^{13}$ electrons /cm$^{2}$\\
 (f) Ref.~\onlinecite{Liao_2015}, DFT (DFPT)\\
 (g) Monte Carlo and DFT (finite differences), acoustic and optical phonons\\
 (h) Monte Carlo and DFT (DFPT) \\
 (i) Kubo-Greenwood, acoustic phonons only, elastic and equipartition approximation 
\par\end{raggedright}
\label{tab:mu_theory} 
\end{table}

\subsection{Available experimental and theoretical results}

\label{sec:Review} 
Given the large number of studies that have been published regarding
the carrier mobility in phosphorene, it is convenient to summarize
in Tables~\ref{tab:mu_exp} and ~\ref{tab:mu_theory} the available
experimental and theoretical results, before commenting on them. A
necessary critical review will follow, as anticipated.

\textit{Experimental information.} As we have mentioned above, interest
in phosphorene has been stimulated by the relatively high room-temperature
carrier mobility measured in black phosphorus: 300 to 1,100 cm$^{2}$V$^{-1}$s$^{-1}$
for electrons and 150 to 1,300 cm$^{2}$V$^{-1}$s$^{-1}$ for holes,
depending on orientation\cite{Akahama_1983}. Information for multi-layers
is limited to the hole mobility, since samples are almost invariably
p-type. Only Cao \textit{et al.}\cite{Cao_2015} have observed ambipolar
behavior in field-effect transistors with mono-, bi-, and tri-layer
channels, finding an electron mobility much smaller than the hole
mobility in all cases. They have also provided the only measurement
for charge-transport properties of monolayers. In all cases, the hole
mobility is strongly anisotropic and shows a strong dependence on
the thickness of the film. Specifically, Cao \textit{et al.}\cite{Cao_2015}
have measured a room-temperature hole mobility of 1, 80, and 1,200
cm$^{2}$V$^{-1}$s$^{-1}$ in mono-, bi-, and tri-layers. Li \textit{et
al.}\cite{Li_2014} have also observed a thickness dependence, reporting
a 300~K hole mobility of around several hundreds cm$^{2}$V$^{-1}$s$^{-1}$
for thick layers, sharply decreasing in layers thinner than 10~nm,
and reaching values as low as 1-10 cm$^{2}$V$^{-1}$s$^{-1}$ for
layers 2-3~nm-thin\cite{Li_2014}. A similar trend has been reported
also by Liu \textit{et al.}\cite{Liu_2014}, with a peak field-effect
hole mobility of 286 cm$^{2}$V$^{-1}$s$^{-1}$ in 5~nm-thick films.
Xia \textit{et al.}\cite{Xia_2014} found a Hall mobility of about
600 cm$^{2}$V$^{-1}$s$^{-1}$ in 15~nm-thick films and of about
400 cm$^{2}$V$^{-1}$s$^{-1}$ in 8~nm-thick films. The general
trend of an increasing mobility in thicker films is also confirmed
by the results of Xiang \textit{et al.}\cite{Xiang_2015}, who have
measured a hole mobility of 214 cm$^{2}$V$^{-1}$s$^{-1}$ in 4.8~nm-thick
films, compared to a Hall mobility of 400 cm$^{2}$V$^{-1}$s$^{-1}$
observed in `few-layers' phosphorene by Gillgren \textit{et al.}\cite{Gillgren_2015}.
As a rare example of information on electron transport, Doganov and
coworkers\cite{Doganov_2015} have reported an electron mobility of
about 106 cm$^{2}$V$^{-1}$s$^{-1}$ in 10~nm-thick films. For holes,
they have measured a value of 189 cm$^{2}$V$^{-1}$s$^{-1}$. Note
that these are field-effect mobilities. Finally, a similar thickness
dependence has been observed also at low temperatures by Tayari and
coworkers\cite{Tayari_2015}, who have considered black-phosphorus
films with thickness in the range of 6.1-to-47~nm and have measured
a maximum mobility of about 600 (below 80~K) and 900 (300 mK) cm$^{2}$V$^{-1}$s$^{-1}$
in their thickest films. In most of these experiments, the highest
field-effect hole mobility has been measured for channels presumably
oriented along the armchair direction ('presumably' only, since this
information is not always given).

\textit{Theoretical results}: Restricting again our attention to room-temperature
data, in monolayers a relatively large hole mobility of about 640-to-700
cm$^{2}$V$^{-1}$s$^{-1}$ has been calculated by Qiao and coworkers
for transport along the armchair direction\cite{Qiao_2014}. Surprisingly,
a huge hole mobility, 10,000-to-26,000 cm$^{2}$V$^{-1}$s$^{-1}$
has been predicted along the heavy-mass zigzag direction\cite{Qiao_2014},
a result allegedly due to an extremely small deformation potential.
For electrons, Qiao \textit{et al.}\cite{Qiao_2014} have calculated
a mobility of 1,100 (armchair) and 80 (zigzag) cm$^{2}$V$^{-1}$s$^{-1}$.
Monte Carlo simulations based on the DFT-calculated band structure
and carrier-phonon scattering rates\cite{Giannozzi_2009} have been
performed by Jin \textit{et al.}\cite{Jin_2016}. They have obtained
a hole mobility of 460 and 90 cm$^{2}$V$^{-1}$s$^{-1}$ for transport
along the armchair and zigzag directions, respectively. For electrons,
these values are, instead, 210 (armchair) and 40 (zigzag) cm$^{2}$V$^{-1}$s$^{-1}$.
They have also predicted a higher mobility in bilayers: 1,020 cm$^{2}$V$^{-1}$s$^{-1}$
(armchair) and 360 cm$^{2}$V$^{-1}$s$^{-1}$ (zigzag) at 300~K
for electrons, 1,610 cm$^{2}$V$^{-1}$s$^{-1}$ (armchair) and 760
cm$^{2}$V$^{-1}$s$^{-1}$ (zigzag) for holes. This enhanced mobility
has been attributed to the smaller hole effective mass in bilayers,
especially along the `flat' $\Gamma-Y$ (zigzag) direction, as indicated
by Qiao \textit{et al.}\cite{Qiao_2014}. We shall later discuss (or, better,
speculate about) possible causes for the observed thickness dependence
of the hole mobility. More recently, Trushkov and Perebeinos\cite{Trushkov_2017}
have calculated the electron mobility as a function of carrier density
and temperature, obtaining values of 625 cm$^{2}$V$^{-1}$s$^{-1}$
(armchair) and 82 cm$^{2}$V$^{-1}$s$^{-1}$ (zigzag) at 300~K and
at an electron density of $10^{13}$ cm$^{-2}$. Rudenko and coworkers\cite{Rudenko_2016}
have obtained similar values of 738 (armchair) and 114 (zigzag) cm$^{2}$V$^{-1}$s$^{-1}$
for electrons, of 292 (armchair) and 157 (zigzag) cm$^{2}$V$^{-1}$s$^{-1}$
for holes. At the lower range of predicted mobility, Liao \textit{et
al.}\cite{Liao_2015} have also used DFT to calculate the band structure
and carrier-phonon matrix elements, obtaining the values of 170 (armchair)
and 50 (zigzag) cm$^{2}$V$^{-1}$s$^{-1}$ for electrons, of 170
(armchair) and 35 (zigzag) cm$^{2}$V$^{-1}$s$^{-1}$ for holes. 

\subsection{Why such a disagreement?}

\label{sec:Critique} 
It is probably premature to compare the experimental and theoretical
results summarized in Tables~\ref{tab:mu_exp} and \ref{tab:mu_theory}.
Indeed, different experimental results may be due to expected deviations
from ideality of the material, such as impurities and defects, resulting
from an immature technology. Moreover, experimental data have been
obtained in supported \textendash{} and often gated \textendash{}
layers, whereas theoretical calculations have considered ideal free-standing
films. A proper discussion of this issue would distract us from the
main focus of this paper. Here we shall only remark that changes of
the phonon spectra may be expected when moving from free-standing
layers to supported and gated materials. This, obviously and in principle,
may affect the carrier mobility. For van der Waals materials such
as graphene, in-plane acoustic modes are left largely unaffected by
interactions with a substrate, even when as strong as coupling with
metals, whereas optical phonons are slightly softened by the dielectric
screening of the metal.\cite{AlTaleb_2016} An excellent review of
phonon dynamics in 2D materials also highlights the 2D nature of the
layer(s) as the major effect that controls the vibrational frequencies\cite{Balandin_2014}.
Moreover, as we have already remarked, most calculations have been performed assuming intrinsic
materials, whereas experiments usually deal with gated layers at a
high carrier density. Given this `circumstantial' evidence, and ignoring
scattering with non-idealities of the substrate/gate (charges and
defects), coupling with hybrid plasmon/substrate-optical modes\cite{Ong_2012},
and considering that acoustic flexural modes \textendash{} indeed
affected by the substrate and the gate \textendash{} do not play any role
in phosphorene, we should not expect gross changes of the electronic-transport
properties, at least in van der Waals materials. However, as discussed below,
phosphorene is not a pure `van der Waals' material. Therefore, it
may couple rather strongly with a substrate or gate insulator. Depending
on the vibrational stiffness (\textit{e.g.}, SiO$_{2}$) or softness
(\textit{e.g.}, HfO$_{2}$) of this layer, the spectrum of the acoustic
and especially of the optical modes of phosphorene will be affected
accordingly. Nevertheless, here we wish to focus on the wide variations
of the theoretical predictions for free-standing layers listed in
Table~\ref{tab:mu_theory}.

Indeed, besides an obvious thickness and orientation dependence, we
have already observed that the wide variations of the experimental
data shown in Table~\ref{tab:mu_exp} may be explained by deviations
from ideality of the material. No such plausible explanation can easily
be found to make sense of the surprisingly wide range of theoretically
predicted values shown in Table~\ref{tab:mu_theory}. This is a disconcerting
observation, since most of the theoretical results have been obtained
by using `first principles' calculations (DFT), often from the same
software packages.

Obviously, even \textit{ab initio} DFT calculations, despite their
elegance and predictive power, exhibit some limitations. For
example, the behavior of the electronic dispersion near the top of
the valence band in mono and multi-layer phosphorene is very flat.
The details of this dispersion (that is, effective mass and velocity)
depend strongly on the exact details of the calculation\cite{Lew_2015,Fukuoka_2015},
so much so that a hole effective mass along the zigzag direction cannot
be defined. Such details may matter only marginally when considering
the overall band structure, but the equilibrium transport properties
are strongly affected by such tiny differences, being extremely sensitive
to variations of the order of the thermal energy. These variations
affect the carrier velocity and the density of states that, in turn,
affects the scattering rates. Moreover, the carrier-phonon matrix
elements obtained from DFT calculations may suffer from errors and
uncertainties related not only to the choice of pseudopotentials and
exchange-correlation functionals, but also to the subtle issue of
dielectric screening. This has been observed to be the case for graphene\cite{Piscanec_2004,Lazzeri_2006,Lazzeri_2008},
resulting in underestimated deformation potentials\cite{Borysenko_2010},
as discussed in Ref.~\onlinecite{Fischetti_2013}. Therefore, different
choices of pseudopotentials or exchange-correlation functionals (local
density approximation \textendash{} LDA, generalized gradient approximation
\textendash{} GGA, or hybrid exchange-correlation functionals), or the
use of GW corrections, will result in different bandstructures and values of the total-energy,
and so in different phonon spectra and electron-phonon matrix elements.

However, it is difficult to see how such uncertainties, as large as
they might be, may result in almost 4 orders of magnitude variations
seen in Table~\ref{tab:mu_theory} for the hole mobility along the
zigzag directions, from 10,000-26,000 cm$^{2}$V$^{-1}$s$^{-1}$
(Ref.~\onlinecite{Qiao_2014}) to the small values of the order unity
that we shall present below.

In our opinion, opinion that is also shered by Nakamura and coworkers\cite{Nakamura_2017}, 
a first {\it major} source of errors is the use of the so-called
Takagi formula\cite{Takagi_1994} to calculate the carrier mobility
for a two-dimensional electron gas (2DEG): 
\begin{equation}
\mu\ =\ \frac{e\hbar^{3}C_{{\rm 2D}}}{k_{{\rm B}}Tm^{\ast}m_{{\rm d}}E_{1}^{2}}\ .
\label{eq:Takagi}
\end{equation}
In this expression, $m^{\ast}$ and $m_{{\rm d}}$ are the conductivity
and density-of-states effective masses, respectively, $C_{2D}$ is
the longitudinal or transverse elastic constant of the 2D materials,
and $E_{1}$ is the so-called `deformation potential'. In this context,
this last all-important quantity is defined as the energy-shift, $\Delta E_{c,v}$,
of the relevant band-edge (conduction for electron transport, valence
for holes), under a relative change, $\Delta a/a_{0}$, of the lattice
constant $a_{0}$, 
\begin{equation}
E_{1}\ =\ a_{0}\frac{\Delta E_{c,v}}{\Delta a}\ .
\label{eq:E1}
\end{equation}
Approximations equivalent to those implied by Eq.~(\ref{eq:Takagi})
have also been used to calculate extremely high values for the carrier
mobility in silicene ($\approx2\times10^{5}$ cm$^{2}$ V$^{-1}$
s$^{-1}$ for both electrons and holes)\cite{Shao_2013} and germanene ($\approx6\times10^{5}$ cm$^{2}$ V$^{-1}$
s$^{-1}$ for both electrons and holes)\cite{Ye_2014}, ignoring coupling to flexural acoustic
modes\cite{Fischetti_2016}. Equation~(\ref{eq:Takagi})
must be used with extreme care. It was originally derived by Takagi
\textit{et al.}\cite{Takagi_1994} in the context of electron transport
in Si inversion layers. It was intended to be used to calculate the
mobility limited by scatterng with acoustic phonons, only one of the many scattering processes that
affect electron transport in those systems, such as intersubband/intervalley
process, scattering with optical phonons, with ionized impurities,
and with surface roughness. Even when taken in this originally limited
context, Eq.~(\ref{eq:Takagi}) is of a semi-empirical nature, since
longitudinal and transverse acoustic phonons are lumped into one single `effective'
mode, with an isotropic deformation potential $E_{1}$, which for Si is typically
in the range of 9-14 eV, fitted to reliably known experimental data.
Indeed, within the framework of the deformation potential theorem\cite{Bardeen_1950},
the electron/acoustic-phonon matrix elements in Si are known to be
anisotropic\cite{Herring_1957}, a property that plays a crucial role
in explaining the electron mobility in strained Si (Ref.~\onlinecite{Fischetti_1996})
and Si inversion layers\cite{Fischetti_1993}. Arbitrarily extending
Eq.~(\ref{eq:Takagi}) to the more general context of 2D crystals
presents severe problems: Scattering with optical modes and intervalley
processes, when present, are neglected by Eq.~(\ref{eq:Takagi}).
Moreover, as we have emphasized, only one phonon mode is considered,
LA or TA, depending on the choice of $E_{1}$ and $C_{2D}$; also
neglected is the anisotropy of the deformation potential, an effect that,
as we have already noted, is extremely important in Si (Refs.~\onlinecite{Herring_1957,Fischetti_1996,Fischetti_1993})
and that has also been shown to be equally important in phosphorene\cite{Liao_2015}.
Finally, the symmetry of the initial and final wavefunctions affects
the magnitude and angular dependence of the carrier-phonon matrix
elements, and these `wavefunction-overlap effects' are also ignored
altogether. Therefore, the results of Qiao \textit{et al.}\cite{Qiao_2014}
should be regarded as no more than extremely optimistic upper bounds.

A second likely source of errors is the use of the `band deformation
potential', Eq.~(\ref{eq:E1}), to approximate the electron-phonon
matrix elements. Even when moving beyond Eq.~(\ref{eq:Takagi}) by
using, for example, the Kubo-Greenwood expression to calculate the
carrier mobility, the shifts of the band-edges under various strain
conditions give only a qualitative approximation for the scattering
matrix elements, since the effects mentioned above \textendash{} mainly,
wavefunction-overlap and angular dependence \textendash{} are still
neglected. Such models have been employed by Trushkov and Perebeinos\cite{Trushkov_2017}
and by Rudenko and coworkers\cite{Rudenko_2016}. With a proper choice
of elastic constants, the latter authors have accounted for both LA
and TA phonons and for two-phonon processes between electrons and
flexural modes. Yet, even in this case, their results, while not quite
so impressive, are still quite large.

The fact that such models employing isotropic deformation potentials
result in an optimistic overestimation of the mobility has been shown
by the work performed by the late professor Dresselhaus' group\cite{Liao_2015}.
Accounting for all modes, for the anisotropy of the matrix elements
\textendash{} due mainly to wavefunction-overlap effects \textendash{}
and accounting also for a non-parabolic band structure, they have
predicted much smaller values. The results that we report here, listed
in Table~\ref{tab:mu_theory}, are even smaller. We shall speculate
below on possible causes for such a disappointing disagreement.

In order to illustrate the importance of the anisotropy of the carrier-phonon
matrix elements, we now present calculations of the electron mobility
in monolayer phosphorene in the simple case of scattering with acoustic
modes only, parabolic bands, linear phonon dispersion, elastic, and
equipartition approximation. The results are sufficiently accurate
to emphasize the importance of the effects ignored by Eqns.~(\ref{eq:Takagi})
and (\ref{eq:E1}). We shall later present in Sec.~\ref{sec:phosphorene}
calculations based on numerically calculated electronic and vibrational
spectra. While more accurate, they are also less transparent. 

\subsection{Electron mobility: A Kubo-Greenwood calculation}

\label{sec:Kubo} 
Considering only scattering with acoustic phonons in the elastic,
equipartition approximation, the electron mobility, $\mu_{\theta}$,
along a crystallographic direction at an angle $\theta$ with respect
to the armchair $\Gamma-X$ direction, is given by the usual Kubo-Greenwood
expression: 
\begin{align*}
\mu_{\theta} & =\ \frac{2e}{n\hbar}\ \int\ \frac{{\rm d}{\bf {k}}}{(2\pi)^{2}}\ \upsilon_{\theta}({\bf {k}})\ \tau_{{\rm p},\theta}({\bf {k}})\ \frac{\partial f({\bf {k}})}{\partial k_{{\rm x}}}\\
 & =\frac{e}{nk_{{\rm B}}T}\ \int\ \frac{{\rm d}{\bf {k}}}{(2\pi)^{2}}\ \upsilon_{\theta}({\bf {k}})^{2}\ \tau_{{\rm p},\theta}({\bf {k}})\ f({\bf {k}})[1-f({\bf {k}})]
\end{align*}
\begin{align}
 & =\frac{2e}{nk_{{\rm B}}T}\ \int_{0}^{\infty}\ \frac{{\rm d}k}{2\pi}k\ \int_{0}^{2\pi}\ \frac{{\rm d}\phi}{2\pi}\ \upsilon_{\theta}(k,\phi)^{2}\ \tau_{{\rm p},\theta}(k,\phi)\nonumber \\
 & \ \ \ \ \times\ f[E(k,\phi)]\{1-f[E(k,\phi)]\}\ ,\label{eq:mu_gen}
\end{align}
where $e$ is the elementary charge, $T$ is the temperature, $k_{{\rm B}}$
is the Boltzmann's constant, $n$ is the carrier density, $E(k,\phi)$
and $\upsilon_{\theta}(k,\phi)$ are the electron energy and group
velocity along the direction $\theta$ of an electron with wave vector
of magnitude $k$ along a direction at an angle $\phi$ with respect
to the armchair direction. Finally, $f(E)$ is the Fermi-Dirac distribution
function.

The momentum relaxation rate, $1/\tau_{{\rm p},\theta}({\bf {k}})$,
along the crystallographic direction $\theta$ results from collisions
with longitudinal (LA) and transverse (TA) acoustic phonons, $1/\tau_{{\rm p},\theta}({\bf {k}})=1/\tau_{{\rm p},\theta}^{{\rm (LA)}}({\bf {k}})+1/\tau_{{\rm p},\theta}^{{\rm (TA)}}({\bf {k}})$.
Using Fermi's golden rule, each rate $1/\tau_{{\rm p},\theta}^{(\eta)}({\bf {k}})$,
with $\eta$ = LA or TA, can be expressed as follows: 
\begin{align}
\frac{1}{\tau_{{\rm p},\theta}^{(\eta)}({\bf {k})}} & \ =\frac{2\pi}{\hbar}\ \int\ \frac{{\rm d}{\bf {k}^{\prime}}}{(2\pi)^{2}}\ \frac{\hbar^{2}DK_{\eta}^{2}({\bf {k},{\bf {k}^{\prime})}}}{2\rho\omega_{{\bf {q}}}^{(\eta)}}\ 2\ \frac{k_{{\rm B}}T}{\hbar\omega_{{\bf {q}}}^{{\rm (\eta)}}}\nonumber \\
 & \times\left[1-\frac{\upsilon_{\theta}({\bf {k}^{\prime})\tau_{{\rm p},\theta}({\bf {k}^{\prime})}}}{\upsilon_{\theta}({\bf {k})\tau_{{\rm p},\theta}({\bf {k})}}}\right]\ \delta[E({\bf {k})-E({\bf {k}^{\prime})]\ .\label{eq:relax_rate}}}
\end{align}

Here $DK_{\eta}({\bf {k},{\bf {k}^{\prime})}}$ is the deformation
potential (implicitly defined by Eq.~(\ref{eq:Defpot}) below) for
initial and final electron wave vectors ${\bf k}$ and ${\bf k}^{\prime}$
and $\omega_{{\bf {q}}}^{{\rm (\eta)}}$ is the frequency of a phonon
of branch $\eta$ and wave vector ${\bf q}$. The phonon-branch index
$\eta$ takes the value LA or TA. Note that the term $2~(k_{{\rm B}}T/(\hbar\omega_{{\bf {q}}}^{{\rm (\eta)}})$
represents the equilibrium occupation number for $\hbar\omega_{{\bf {q}}}^{{\rm (\eta)}}<<k_{{\rm B}}T$,
the additional factor of 2 accounting for emission and absorption
processes.

For simplicity, we approximate the low-energy range of the phosphorene
band structure, calculated as we shall discuss in Sec.~\ref{sec:DFT}
below, with an elliptical parabolic dispersion with a light `armchair'
mass, $m_{{\rm x}}$ = 0.14 $m_{0}$ (where $m_{0}$ is the free electron
mass) and a heavy `zigzag' mass, $m_{{\rm y}}$ = 1.2 $m_{0}$. These
values have been obtained from the DFT calculations discussed in Sec.~\ref{sec:phosphorene}
below. 

\begin{figure}[tb]
\includegraphics[width=7cm]{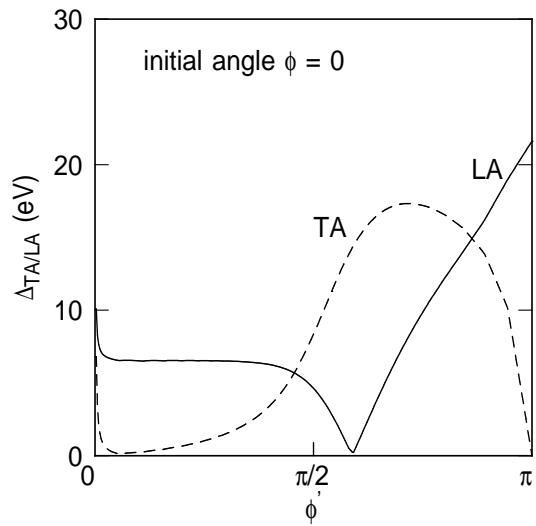} \caption{Acoustic deformation potentials, $\Delta_{{\rm LA}}$ and $\Delta_{{\rm TA}}$,
as of the final scattering angle $\phi^{\prime}$ (with
respect to the armchair direction) for an initial angle $\phi$=0,
that is, for an initial ${\bf k}$ state along the armchair direction.}
\label{fig:P_Delta} 
\end{figure}

The `deformation potential' $DK_{\eta}({\bf {k},{\bf {k}^{\prime})}}$,
has also been calculated from DFT, following Ref.~\onlinecite{Vandenberghe_2015}
as discussed in Sec.~\ref{sec:el-ph} below. The deformation potential exhibits the typical
behavior of the electron/acoustic-phonon matrix elements, vanishing
at long wavelength (that is, at small $q=|{\bf k}-{\bf k}^{\prime}|$).
We find it convenient to calculate $DK_{\eta}({\bf {k},{\bf {k}^{\prime})}}$ on the energy-conserving shell
at an arbitrary (small) electron kinetic energy, $E_{0}$, and express
its complicated but extremely important angular dependence as: 
\begin{equation}
DK_{\eta}({\bf {k}},{\bf {k}^{\prime}})\ =\ (DK)_{0,\eta}\left(\frac{E}{E_{0}}\right)^{1/2}g_{\eta}(\phi,\phi^{\prime})\ ,\label{eq:DK}
\end{equation}
tabulating the functions $g_{LA}(\phi,\phi^{\prime})$ and $g_{TA}(\phi,\phi^{\prime})$
obtained numerically. Figure~(\ref{fig:P_Delta}) shows the quantities
$\Delta_{{\rm LA}}(\phi^{\prime})$ and $\Delta_{{\rm TA}}(\phi^{\prime})$,
defined as $\Delta_{\eta}(\phi^{\prime})=DK_{\eta}({\bf {k}},{\bf {k}^{\prime}})/q$,
where $\phi^{\prime}$ is the angle between the armchair direction
and the direction of the final state ${\bf k}^{\prime}$ for the particular
case of an initial state, ${\bf k}$, along the armchair direction
($\phi$ = 0).

\begin{figure}[tb]
\includegraphics[width=7cm]{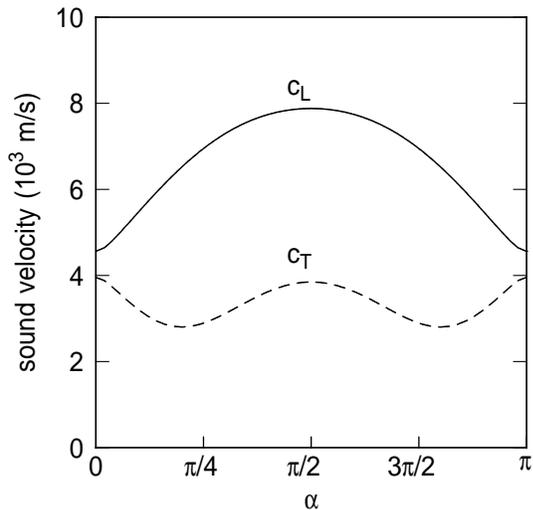} \caption{Longitudinal, $c_{{\rm L}}$, and transverse, $c_{{\rm T}}$, sound
velocity as functions of the angle $\alpha$ between the phonon wave
vector and the $\Gamma-X$ symmetry line.}
\label{fig:P_csound} 
\end{figure}

\begin{figure}[tb]
\includegraphics[width=7cm]{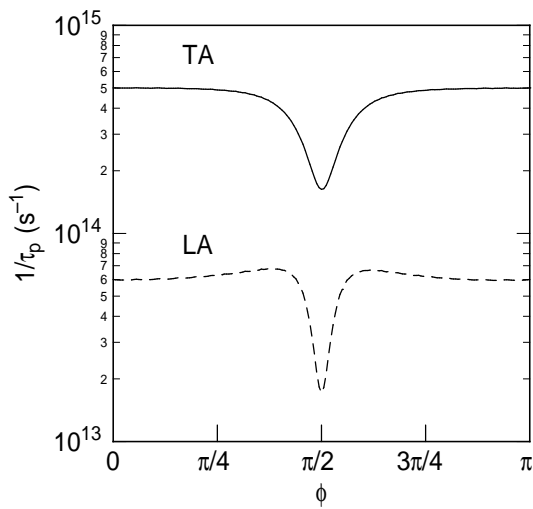} \caption{Momentum relaxation rate for electron scattering with acoustic phonons,
as a function of the angle $\phi$ between the ${\bf k}$ vector and
the armchair direction. Note that in the parabolic-band and elastic-equipartition
approximations, the momentum relaxation rate does not depend on the
electron kinetic energy.}
\label{fig:P_relax} 
\end{figure}

We also use the DFT results shown in Fig.~\ref{fig:El_ph_bands}b
to approximate the frequency of the acoustic phonons in the neighborhood
of the $\Gamma$ point with linear dispersions via a longitudinal
and a transverse sound velocity, $c_{\eta}(\alpha)$, that depends
on the angle $\alpha$ between the $\Gamma-X$ symmetry line and the
phonon wave vector ${\bf q}$, that is, $\omega_{{\bf q}}^{(\eta)}\approx c_{\eta}(\alpha)q$.
Figure~\ref{fig:P_csound} shows the angular dependence of the sound
velocity along the longitudinal (armchair) and transverse (zigzag)
directions, whereas in Fig.~\ref{fig:P_relax} we show the angular
dependence of the momentum relaxation rate. 

With these approximations, the momentum relaxation rate $1/\tau_{{\rm p},\theta}^{(\eta)}(k,\phi)$
becomes independent on the magnitude of the wave vector (that is,
of energy): 
\begin{align}
\frac{1}{\tau_{{\rm p},\theta}^{(\eta)}(k,\phi)}\  & =\ \frac{(DK)_{0,\eta}^{2}k_{{\rm B}}T}{\hbar^{3}\rho}\ \frac{E(k,\phi)}{E_{0}}\ \int_{0}^{2\pi}\frac{{\rm d}\phi^{\prime}}{2\pi}\nonumber \\
 & \hspace*{-1.75cm}\times\ \frac{g_{\eta}^{2}(\phi,\phi^{\prime})}{c_{\eta}^{2}(\alpha)q^{2}(\cos^{2}\phi/m_{{\rm x}}+\sin^{2}\phi/m_{{\rm y}})}\ \left[1-\frac{{\bm{\upsilon}}({\bf {k}^{\prime})\cdot{\bm{\upsilon}}({\bf {k})}}}{\upsilon({\bf {k}}^{\prime})\upsilon({\bf {k}})}\right]\ .\label{eq:relax_rate_1}
\end{align}
Note that in this equation we have approximated the `velocity-loss'
factor (given by the last factor in square brackets inside the integrand
in Eq.~(\ref{eq:relax_rate})) with the expression suggested in Refs.~\onlinecite{Gunst_2016}
and \onlinecite{Liao_2015}, an expression that depends only on the
angle between the initial and final group velocities, ${\bm{\upsilon}}({\bf {k})}$
and ${\bm{\upsilon}}({\bf {k}^{\prime})}$, not on their magnitude,
thanks to the elastic approximation. Therefore, in Eq.~(\ref{eq:mu_gen})
the integration over $k$ simplifies and the electron mobility can
be simplified to the following expression that does not depend on
the carrier density $n$ (or Fermi energy\cite{note_3}): 
\begin{align}
\mu_{\theta} & =\frac{e}{\pi m_{{\rm d}}}\int_{0}^{2\pi}{\rm d}\phi\nonumber \\
 & \left(\frac{\cos\phi\cos\theta/m_{{\rm x}}+\sin\phi\sin\theta/m_{{\rm y}}}{\cos^{2}\phi/m_{{\rm x}}+\sin^{2}\phi_{{\rm y}}/m_{{\rm y}}}\right)^{2}\ \tau_{{\rm p},\theta}(\phi)\ .\label{eq:mu_gen_1}
\end{align} \\

\begin{figure}[tb]
\includegraphics[width=7.5cm]{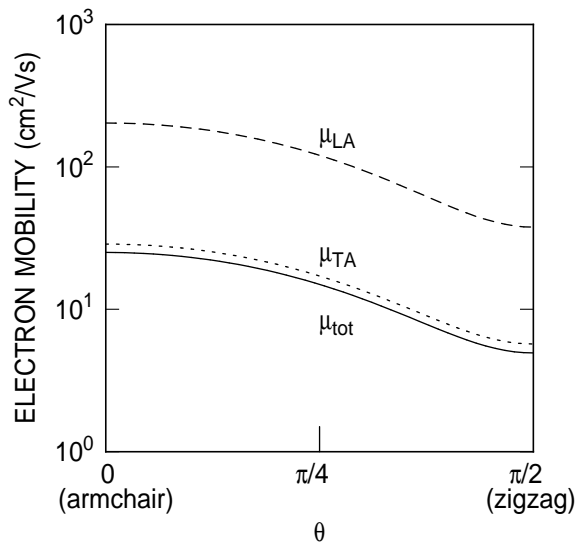} \caption{Electron mobility in monolayer phosphorene at 300~K as a function
of the angle $\theta$ between the transport direction and the armchair
$\Gamma-X$ symmetry line. The LA-phonon- and TA-phonon-limited mobilities
are shown separately. These results have been obtained using the parabolic-band
and elastic-equipartition approximations and ignoring scattering with
optical phonons.}
\label{fig:P_mu_angle} 
\end{figure}

The electron mobility obtained from Eq.~(\ref{eq:mu_gen_1}) is shown
in Fig.~\ref{fig:P_mu_angle}. Despite the relatively simple models
we have used so far, the results are in excellent qualitative agreement,
and even reasonable quantitative agreement, with our more accurate
full-band Monte Carlo results presented in Sec.~\ref{sec:P-mono}
below. Yet, our results are significantly different from those reported
in Refs.~\onlinecite{Jin_2016} and \onlinecite{Liao_2015}.

Such a disagreement is not too surprising when considering the hole
mobility: The `flatness' of the dispersion of the highest-energy valence
band along the zigzag direction seems to depend very strongly on the
computational method used, as already pointed out by Lew Van Yoon
and coworkers\cite{Lew_2015}. This uncertainty affects the velocity
and density of states of thermal holes, so that, sadly, results vastly
different should not come as a surprise.

More disconcerting is the situation regarding the electron mobility.
We can only speculate on possible physical and numerical causes of
the difference between our results and those reported in Refs.~\onlinecite{Jin_2016}
and \onlinecite{Liao_2015}. In Ref.~\onlinecite{Liao_2015} the local density 
approximation (LDA)\cite{LDA} is used for the exchange correlation functional, as
opposed to the Perdew-Burke-Enzerhof generalized-gradient approximation (GGA-PBE)\cite{Perdew_1996} 
we have used. This likely yields a difference in both electronic and phononic structure, which may
at least in part, account for the observed difference. 
We preferred PBE over LDA since it is known to yield a better approximation for the 
exchange-correlation functional than LDA. This observation, coupled to possible and 
likely different phonon polarization vectors, may be the cause of the much larger deformation
potential we have obtained for the electron/TA-phonon interaction in particular.
Indeed, as we shall emphasize below, great care must be taken, by using
large supercells for the phonon calculations, to avoid `imaginary' frequencies for the low-energy
vibrational modes and, even when obtaining a `correct' vibrational
spectrum, the resulting eigenvectors are only known with relatively small
accuracy. Moreover, the calculation of the mobility requires numerical
integrations over the Brillouin zone and we have found that the results
require a fine discretization to perform such
integrations. Specifically, as we shall see below, we have used a
very fine mesh around the $\Gamma$ symmetry point, equivalent to
$145\times205$ ${\bf k}$-points in the first quadrant. On the contrary,
a much coarser mesh, presumably $12\times12$ points in the entire
BZ, has been employed in Ref.~\onlinecite{Jin_2016}.
Despite the quantitative disagreement with the \textit{ab initio}
results presented in Refs.~\onlinecite{Jin_2016} and \onlinecite{Liao_2015},
we confirm their main conclusion: Besides the obviously strong anisotropy
of the mobility, due mainly to the anisotropy of the conduction bands,
these results show how important the matrix-element anisotropy really
is, an effect that explains the large values calculated by Qiao \textit{et
al.}\cite{Qiao_2014} and also those reported in Refs.~\onlinecite{Trushkov_2017}
and \onlinecite{Rudenko_2016}. 

\section{First-principles physical models and numerical methods}

\label{sec:Methods} 
Having used a simplified model to draw some early conclusions, we now
consider more accurate but less transparent {\it ab initio} methods.

\subsection{Band structure and phonon spectrum}

\label{sec:DFT} 

For the calculation of the band structure of the systems considered
here, we have primarily used the Vienna \textit{{Ab~initio}} Simulation Package
(VASP)\cite{VASP1,VASP2,VASP3,VASP4} with the Perdew-Burke-Enzerhoff
generalized-gradient approximation (GGA-PBE) for the exchange-correlation
functional\cite{Perdew_1996} and a projector augmented wave (PAW)\cite{Blochl_1994}
pseudopotential. A vacuum `padding' of 20 \AA\ was
used in constructing the unit-cell to obtain free-standing layers 
and avoid interaction with periodic images. Initially, we have always performed
structural optimization by minimizing the total energy in
order to determine the lattice constants and ionic positions. 
VASP also handles van der Waals (vdW) interactions, important in multilayer systems,
with Grimme's model\cite{Grimme_2004}. We have chosen the `optPBE'
functionals\cite{optPBE} among the various other vdW-corrected functionals\cite{optB88,vdW-DF} for the bilayers.
We shall further discuss this issue below. 
The phonon spectra have been obtained using the PHONOPY computer program\cite{PHONOPY} which
calculates the force constants using small-displacement method, using an interface to VASP 
to obtain the atomic forces.
In our calculations, we have found that a supercell of size of at least $8\times8\times1$ unit-cells
is required to avoid unphysical imaginary frequencies for low-energy
acoustic phonons, especially for the flexural out-of-plane (ZA) modes. 

In light of the large discrepancies seen in literature when using different methods, 
we have decided to verify the results obtained from VASP using a different implementation of DFT, 
the Quantum Espresso (QE)\cite{Giannozzi_2009} DFT package, so that we can independently 
calculate the atomic configuration and electronic structure for monolayer phosphorene using ultrasoft pseudopotentials, and the PBE-GGA exchange-correlation approximation. 
The phonon spectra have been obtained by using QE, which yields the dynamical matrix using Density Functional Perturbation Theory (DFPT)\cite{Baroni_2001}, as opposed to the small-displacements method of PHONOPY.
The phonon spectra obtained from QE are calculated on a coarse $\bold{q}$-points grid and interpolation 
to a fine grid, as required for the proper estimation of the scattering rates, is performed with minimal loss of
accuracy by using maximally localized Wannier functions as implemented in the EPW\cite{epw_Giustino,epw_Giustino2} package.
In QE, the $\bold{k}$-point grid used in the self-consistent calculation of the electronic structure must 
be a multiple of the $\bold{q}$-point grid to obtain an accurate phonon dispersion. 

For both the VASP and QE methods, the self-consistent calculations should be 
performed using a large cutoff energy to avoid negative frequencies, especially for the ZA-phonons.
Distinguishing among longitudinal and transverse, and between acoustic
and optical modes, has proven difficult or impossible (and even unphysical), especially
for the many optical phonons in bilayers.
The computational parameters we used are shown in Table~\ref{tab:Comput_dft}.
It was observed that, even though the methods differ significantly in their approach, both VASP and Quantum Espresso yield matching crystal structure, band structure, and phonon dispersion for monolayer phosphorene.

\begin{table}
\centering
\caption{Computational parameters for DFT calculations.}
\label{tab:Comput_dft}
\begin{ruledtabular}
\begin{tabular}{ccc}
Parameters & VASP & QE \tabularnewline
\hline 
$E_\mathrm{k}$ cutoff & 500~eV  &  50~Ry   \tabularnewline
Ionic minimization threshold & $10^{-6}$~eV & $10^{-6}$~Ry\tabularnewline
SCF threshold &$10^{-8}$~eV & $10^{-10}$~Ry\tabularnewline
$\bold{k}$ points mesh& $11\times11\times1$&$12\times12\times1$ \tabularnewline


\hline
EPW meshes & coarse & fine  \tabularnewline
\hline 
$\bold{k}$ points mesh& $12\times12\times1$  & $50\times50\times1$  \tabularnewline
$\bold{q}$ points mesh& $6\times6\times1$& $50\times50\times1$\tabularnewline
\end{tabular}
\end{ruledtabular}

$E_\mathrm{k}$ : Kinetic Energy\\
SCF : Self consistent field

\end{table}

\subsection{Carrier-phonon interaction}

\label{sec:el-ph} 
We have treated the electron-phonon interaction following the general
theory developed in Refs.~\onlinecite{Baroni_2001,Giustino_2017,Borysenko_2010,Vandenberghe_2015}.
The matrix elements for the electron-phonon interaction can be expressed as: 
\begin{multline}
\hspace*{-0.5cm}\langle{\bf k}^{\prime}n^{\prime}|V_{{\bf q}}^{(\eta)}|{\bf k}n\rangle=\left\{ \begin{array}{cc}
n_{{\bf q}}^{(\eta)1/2}\\
(1+n_{{\bf q}}^{(\eta)})^{1/2}
\end{array}\right\} \sum_{l,\gamma}\left(\frac{\hbar}{2N_{{\rm c}}M_{\gamma}\omega_{{\bf q}}^{(\eta)}}\right)^{1/2}\\
\times\ e^{{\rm i}{\bf q}\cdot{\bf R}_{l\gamma}}\ {\widehat{{\bf e}}}_{{\bf q},\gamma}^{(\eta)}\cdot\int_{\Omega}\ {\rm d}{\bf r}\ \psi_{{\bf {k}'n^{\prime}}}({\bf r})^{\ast}\ \frac{\partial U({\bf r})}{\partial{\bf R}_{0,\gamma}}\ \psi_{{\bf {k}n}}({\bf r})\ ,\label{eq:epmat}
\end{multline}
where $N_{{\rm c}}$ is the number of cells, $M_{\gamma}$ the mass
of ion $\gamma$ in each cell, $\Omega$ is the volume of the crystal,
the index $l$ labels the cells, ${\bf R}_{l\gamma}$ the equilibrium
position of ion $\gamma$ in cell $l$, ${\bf k}$, ${\bf k}^{\prime}$
and $n$,$n^{\prime}$ are the wave vectors and band indices of the
initial and final electronic states, respectively, $\psi_{{\bf {k}n}}({\bf r})$
are the associated Bloch wavefunctions, $\omega_{{\bf q}}^{(\eta)}$
is the phonon frequency and ${\widehat{{\bf e}}}_{{\bf q},\gamma}^{(\eta)}$
the unit displacement vector of ion $\gamma$ for a phonon of branch
$\eta$ and wave vector ${\bf q}={\bf k}^{\prime}-{\bf k}$. Here, $\bf r$, ${\widehat{{\bf e}}}_{{\bf q},\gamma}^{(\eta)}$ and ${\bf R}_{l\gamma}$
are 3D-vectors while ${\bf k}$, ${\bf k}^{\prime}$ and ${\bf q}$ are 2D-vectors. The quantity
$n_{{\bf q}}^{(\eta)}$ is the Bose-Einstein phonon occupation number
(obtained after having implicitly traced-out the phonon modes, assumed
to be at equilibrium) for phonons of branch $\eta$ and wave vector
${\bf q}$. The upper (lower) term within the curly brackets applies
to phonon absorption (emission) processes. Finally, the term $\partial U({\bf r})/\partial{\bf R}_{l,\gamma}$
represents the change of total energy of the lattice under a shift
$\delta{\bf R}_{l,\gamma}$ of the position of ion $\gamma$ in cell
$l$. Following from the two methods used in the evaluation of the electron and, in particular, the phonon spectra, we have evaluated (Eq.~(\ref{eq:epmat})) using two distinct methods.

In our primary method, based on the VASP and PHONOPY packages, the term is approximated by 
identifying $U({\bf r})$ with the Hartree component of the Kohn-Sham Hamiltonian, $U_{{\rm H}}({\bf r})$, 
and is evaluated using finite differences.
Of course, the periodicity of the lattice implies that this quantity does not depend on the cell-index $l$.
We have numerically evaluated this term following Ref.~\onlinecite{Vandenberghe_2015} using VASP.
In our second method, the term is evaluated entirely within the DFPT formalism using the Quantum Espresso 
and EPW packages, where $U({\bf r})$ now accounts for both the Hartree and the exchange/correlation 
components of the potential. 
Once again, we have used both methods for monolayer phosporene to verify our results,
and we have found that the obtained results are in excellent agreement.
Perhaps surprisingly, we must therefore conclude that our results are not very sensitive to the actual method used, provided that sufficient care is taken to use very low tolerances and fine grids in order to capture all the relevant physics.

We shall also make frequent reference to the `deformation potential',
$DK_{\eta}({\bf k}n,{\bf k}^{\prime}n^{\prime})$, a quantity defined
implicitly by: 
\begin{multline}
\langle{\bf k}^{\prime}n^{\prime}|V_{{\bf q}}^{(\eta)}|{\bf k}n\rangle\ =\ \left\{ \begin{array}{cc}
 n_{{\bf q}}^{(\eta)1/2}\\
(1+n_{{\bf q}}^{(\eta)})^{1/2}
\end{array}\right\} \\
\times DK_{\eta}({\bf k}n,{\bf k}^{\prime}n^{\prime})\ \left(\frac{\hbar}{2M_{cell}\omega_{{\bf k}-{\bf k}^{\prime}}^{(\eta)}}\right)^{1/2}\ ,\label{eq:Defpot}
\end{multline}
having gone to infinite-area normalization, where $M_{cell}$ is the total mass of the supercell. 
The scattering rate of an electron in band (or subband) $n$ and 
in-plane wave vector ${\bf k}$ due to a perturbation potential $V_{{\bf q}}^{(\eta)}$
can now be expressed as an integral only over 2D states as follows: 
\begin{multline}
\frac{1}{\tau^{(\eta)}({\bf k},n)}\ =\ \frac{2\pi}{\hbar}\ \sum_{n^{\prime}}\int {\bf dk}^{\prime}\ |\langle{\bf k}^{\prime}n^{\prime}|V_{{\bf k}-{\bf k}^{\prime}}^{(\eta)}|{\bf k}n\rangle|^{2}\\
\times\ \delta[E_{n}({\bf k})-E_{n'}({\bf k}')\pm\hbar\omega_{{\bf k}-{\bf k}^{\prime}}^{(\eta)}]\ ,\label{eq:tau2D}
\end{multline}
where $E_{n}({\bf k})$ is the energy of an electron or hole with
wave vector ${\bf k}$ in band $n$.

\subsection{Monte Carlo simulations}

\label{sec:Monte Carlo} 
In order to calculate electronic transport properties employing the
first-principles information we have discussed, we have followed the
well-known `full-band Monte Carlo' method to solve numerically the
Boltzmann's transport equation for a two-dimensional electron gas. Such
a method, described, for example, in Refs.~\onlinecite{Jin_2016},
\onlinecite{Borysenko_2010}, or \onlinecite{Fischetti_2013}, requires
the discretization of the BZ into elements centered at points ${\bf k}_{j}$.
The energy $E_{jn}=E_{n}({\bf k}_{j})$ and gradients $\nabla E_{jn}=\nabla_{{\bf k}}E_{n}({\bf k}_{j})$
for each band $n$ are computed, stored in tables, and used to interpolate
the carrier energy and group velocity. Using the Gilat-Raubenheimer
algorithm\cite{Gilat_1966} in two-dimensions\cite{Fischetti_2011},
the same discretization in reciprocal space is used to evaluate numerically
the carrier-phonon scattering rates, Eq.~(\ref{eq:tau2D}), as a
sum over energy-conserving mesh elements in the BZ: 
\begin{multline}
\frac{1}{\tau^{(\eta)}({\bf k},n)}
\approx\ \frac{2\pi}{\hbar}\ \sideset{}{}\sum_{jn^{\prime}}\,\Omega_{xy}
\left|\langle{\bf k}_{j}n^{\prime}|V_{{\bf k }-{\bf k }_{j}}^{(\eta)}|{\bf k}n\rangle\right|^{2}\ \\
\times\frac{1}{(2\pi)^{2}}\ \frac{L(w_{jn'})}{|\nabla E_{jn'}|}\ .\label{eq:tau2D_GR}
\end{multline}
where $\Omega_{xy}$ is the area in the $(x,y)$ plane.
In the notation of Eqns.~(7) and (8) of Ref.~\onlinecite{Fischetti_2011},
here $[1/(2\pi)^{2}]\ L(w_{jn})/|\nabla E_{jn}|$ is the density of
states on band $n$ in the $j^{th}$ element with energy $E_{jn'}=E_{n}({\bf k})\pm\hbar\omega_{{\bf k}-{\bf k}_{j}}^{(\eta)}$
and gradient $\nabla E_{jn'}$ at the center of the element, ${\bf k}_{j}$.
Details about the discretization depend on the particular crystal
structure considered and will be given below. 
In all cases, energy conservation is numerically maintained within a 
root-mean-square error of less than 1~meV.

We have employed a synchronous ensemble Monte Carlo method, in light
of its possible extension to the study of transients and inhomogeneous
cases, although such an extension is not required here. The ensemble
typically consists of 500-to-1,000 `particles', with a time step of
0.2 fs, and followed until steady-state is reached in the uniform
electric field we consider. Usually steady-state is only reached after
several hundreds of ps at high fields, or even ns at low fields. 
We have also assumed a non-degenerate
situation in order to avoid complications originating from Pauli's
exclusion principle. Therefore, our study is constrained to the low-density
limit. Moreover, we have obtained the low-field carrier mobility using Einstein relation by
calculating the diffusion constant $D_{\theta}$ along the direction
$\theta$, a calculation that is less affected by stochastic noise when
the drift velocity is much smaller than the thermal velocity\cite{Jacoboni_1983}.
The diffusion constant is evaluated from the Monte Carlo estimator:
\begin{equation}
D_{\theta}\ =\ \frac{1}{2}\frac{{\rm d}}{{\rm d}t}\left\langle (x_{\theta}-\langle x_{\theta}\rangle)^{2}\right\rangle \ ,\label{eq:diff_const}
\end{equation}
where $\langle x_{\theta}\rangle$ is the time-dependent ensemble-average position
along the direction $\theta$ of electrons initially at the origin,
${\bf r}=0$, diffusing in the absence of an electric field.

\section{Electronic transport}

\label{sec:phosphorene} 

\begin{figure*}
    \centering
    \subfloat{\includegraphics{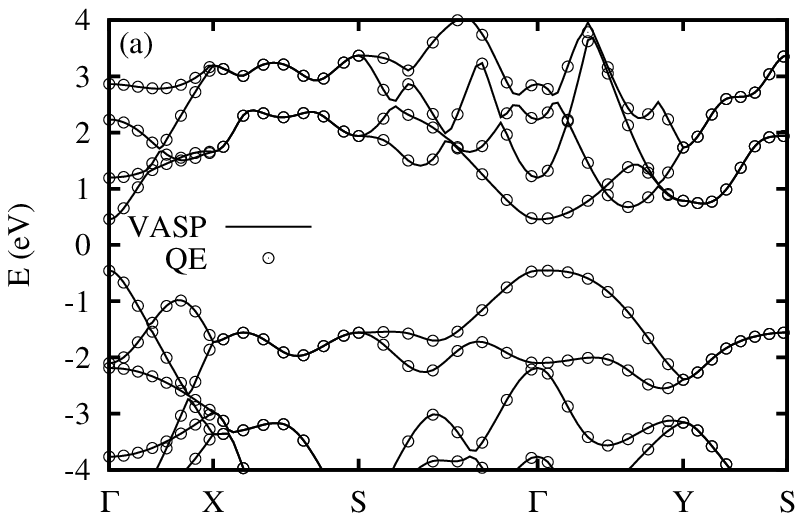}}
    \hfill
    \subfloat{\includegraphics{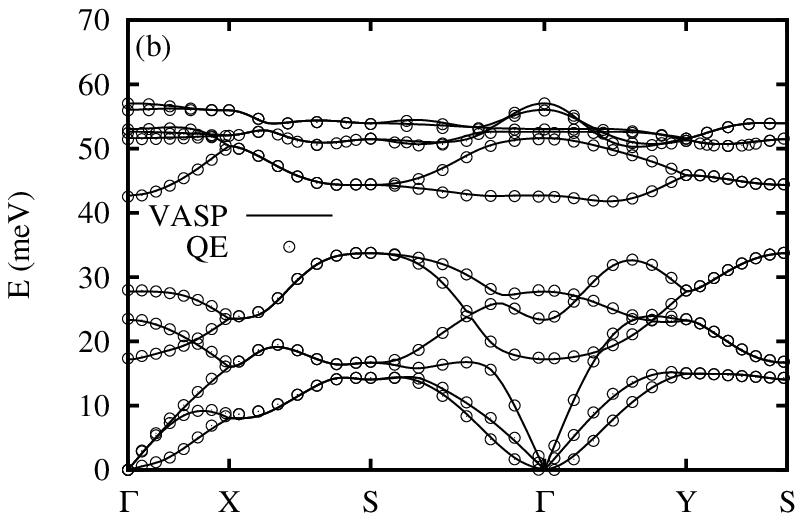}}
    \vfill
    \subfloat{\includegraphics{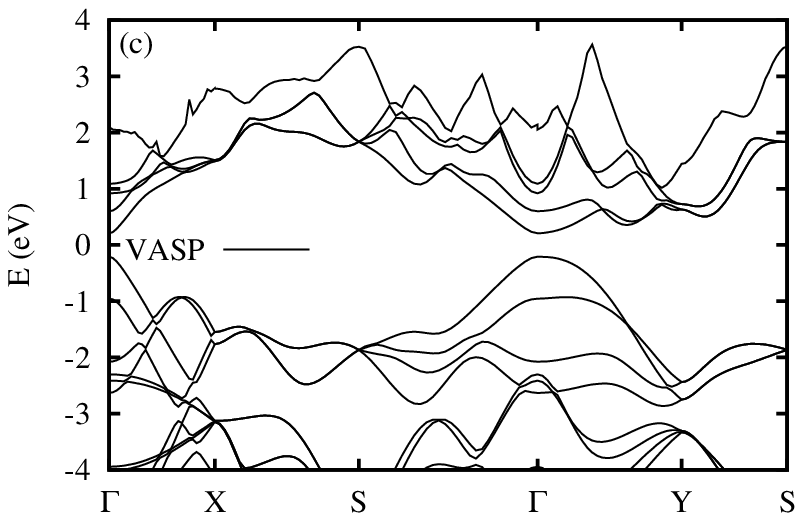}}
    \hfill
    \subfloat{\includegraphics{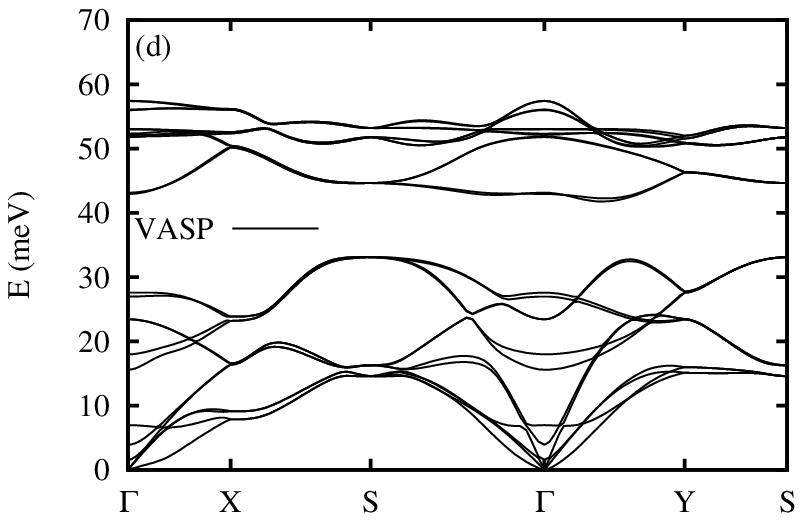}}
    \caption{The calculated band structure (left frames, (a) and (c)) and phonon dispersion (right frames, (b) and (d)), 
             for monolayer phosphorene in (a) and (b), for bilayer phosphorene in (c) and (d)  }
    \label{fig:El_ph_bands}
\end{figure*}

The lattice constants obtained from the structure-relaxation procedure
described above are 4.62~\AA\ and 3.30~\AA\ for monolayer and 4.51~\AA\ and
3.30~\AA\ for bilayer phosphorene, both for the VASP and QE methods, 
and in agreement with the trend reported in Ref.~\onlinecite{Qiao_2014}.
For the bilayer, we obtain a van der Waals gap of 3.20~\AA, a value
that is in excellent agreement with the value calculated by Qiao \textit{et al.}\cite{Qiao_2014}.
For the tabulation and interpolation of the band structure and phonon
spectra obtained using VASP and required by the Monte Carlo simulations, 
the band structure has been tabulated over two nested meshes in the first quadrant of
the $\pi/a\times\pi/b$ Brillouin zone: A `coarse' mesh consisting
of $49\times68$ elements of size $\Delta k_{x}\approx\Delta k_{y}\approx1.38\times10^{8}$/m,
and a `fine' mesh \textendash{} of $44\times205$ elements, of size
$\Delta k_{x}\approx\Delta k_{y}\approx4.63\times10^{7}$/m \textendash{}
in a rectangle with sides of length $0.3\pi/a$ and $\pi/b$ with
the `lower left corner' at the $\Gamma$ point. We found empirically
that the use of such a very fine mesh is necessary to account correctly
for the anisotropy and strong non-parabolicity of the electronic dispersion
around the center of the BZ and in proximity of the local energy minimum along
the symmetry line $Q$.
The phonon spectra and carrier-phonon matrix
elements have also been calculated and tabulated over the same meshes.
However, when using the small displacement, finite differences method, 
$DK_{\eta}({\bf k}n,{\bf k}^{\prime}n^{\prime})$ has been
stored for ${\bf k}^{\prime}$-points on nested meshes with elements
of the same size, but covering the entire BZ, whereas symmetry permits
the tabulation for ${\bf k}$-points on the irreducible wedge (\textit{i.e.},
the first quadrant of the BZ). The size of the tabulated data and
the computation time required to calculate them has been optimized
by considering only `energy conserving' matrix elements; that is:
the deformation potential $DK_{\eta}({\bf k}n,{\bf k}^{\prime}n^{\prime})$
has been calculated and stored only when 
$|E_{n}({\bf k})-E_{n^{\prime}}({\bf k}^{\prime})|\le\hbar\omega_{{\bf k}-{\bf k}^{\prime}}^{(\eta)}$.
{\it We should stress that initial attempts to calculate the deformation potentials over a coarser
mesh have resulted in an inaccurate treatment of the scattering-angle dependence of the deformation potentials
and in an overestimation of the carrier mobility.} 
For the DFPT method, using the QE and EPW packages, we have calculated the carrier-phonon matrix elements
using a coarse $\bold{q}$ and $\bold{k}$ mesh and 
interpolated them to a fine $\bold{q}$ and $\bold{k}$ mesh using 
maximally localized Wannier functions. The coarse and fine mesh sizes used are shown in Table~\ref{tab:Comput_dft}.
During the Monte Carlo simulation, the carrier-phonon matrix element is further interpolated 
on our band structure meshes using bi-linear interpolation.

Given the high computational cost of calculating the deformation potentials,
we have tabulated them only for the first bands (highest-energy valence
band and lowest-energy conduction band) for monolayer phosphorene,
since the next lowest-energy bands are separated by more than 0.7~eV
from the band edges. Therefore, the higher-energy bands we have ignored
are not expected to play any role in the limited set of cases we have
considered and may be ignored. On the other hand, for bilayers, we
had to account for the next lowest-energy conduction band, since
the first and second conduction bands are degenerate at the local
minimum along the point $Q$, so that the second band is expected
to be populated by electrons at the highest fields we have considered.

The electronic band structure and phonon dispersion plotted along
the symmetry points for mono- and bi-layer phosphorene from VASP and QE are 
shown in Fig.~\ref{fig:El_ph_bands}.
DFT calculations are known to underestimate the band
gap. Indeed, the values we obtain for monolayer and bilayer phosphorene
are 0.90~eV and 0.41~eV, lower than experimental values\cite{Qiao_2014,Tran_2014,Zhang_2014}
and also smaller than the results obtained from GW calculations for
mono-layers, $\approx$ 2.0 eV (Ref.~\onlinecite{Fei_2014a}). However,
besides the obvious and unavoidable impact on the effective mass (usually
slightly underestimated when the band gap is also underestimated),
the underestimated band gaps are not expected to significantly affect the transport properties
of interest, since interband transitions between the valence and conduction
band are not included in our calculations. 

\begin{figure}[tb]
\includegraphics[width=8.50cm]{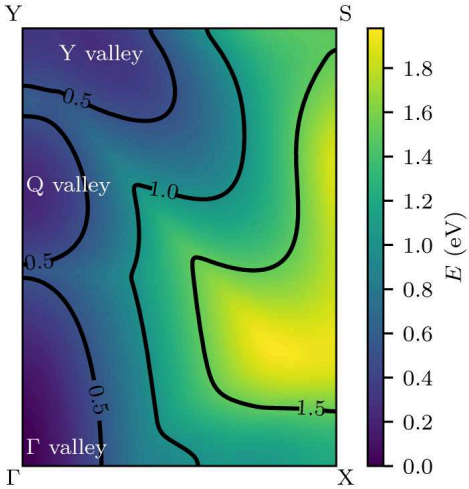}
\caption{Contour plot in the irreducible wedge of the Brillouin zone of the energy of the first conduction band in monolayer
         phosphorene.}
\label{fig:P_mono_contour} 
\end{figure}

The contour plot of the
first conduction band is shown in Fig.~\ref{fig:P_mono_contour} only
for monolayers (bilayers look qualitatively very similar). It indicates
the presence of two satellite valleys, one with a minimum at the symmetry
point $Q$, and a second valley, called the $Y$-valley minimum here,
with its minimum in proximity to a point close to the $Y-S$ symmetry
line. For monolayers, the energy separation between the $\Gamma$
and the $Q$-valley minima, $\Delta E_{\Gamma Q}$ is about 0.21~eV,
whereas the energy separation between the $\Gamma$ and the $Y$-valley
minima $\Delta E_{\Gamma Y}$ is about 0.27~eV. The electron effective
masses in the nearly-isotropic $Q$-valley are 0.25 and 0.30 $m_{0}$
along the armchair and zigzag directions, respectively, as obtained
from the band structure shown in Fig.~\ref{fig:El_ph_bands}a. For
bilayers, Fig.~\ref{fig:El_ph_bands}c, we find similar values, both for the valley energies, 
$\Delta E_{\Gamma Q}$= 0.14~eV and $\Delta E_{\Gamma Y}$ = 0.29~eV, and for the electron
effective masses in the $Q$-valley, 0.25 and 0.32 $m_{0}$ along
the armchair and zigzag directions, respectively.

Comparing the phonon spectra for monolayers, shown in Fig.~\ref{fig:El_ph_bands}b,
to those calculated for bilayers, shown in Fig.~\ref{fig:El_ph_bands}d, we
note the presence of low-energy optical modes for bilayers.
The existence of such low-frequency soft modes is related to the weakness
of the inter-layer coupling (discussed more at length in Sec.~\ref{sec:P-thickness}
below) and also because of the heavy mass of entire unit cells in
different layers oscillating out-of-phase in either the in-plane (LO, TO) or out-of-plane
(ZO) direction. The presence of low-energy optical modes in bilayer phosphorene
was also recently discussed by Xin-Hu \textit{et al.}\cite{Xin-Hu_2016},
and has been already discussed in other multilayer systems, such as
bilayer graphene\cite{Yan_2008,Park_2008}. Moreover, coupling of
these modes with electrons was also shown to be strong\cite{Park_2008}.
We shall discuss below how these modes do indeed affect electronic
transport in bilayer phosphorene.

\subsection{Phosphorene monolayers}

\label{sec:P-mono} 

\begin{figure}[tb]
\hbox{\includegraphics{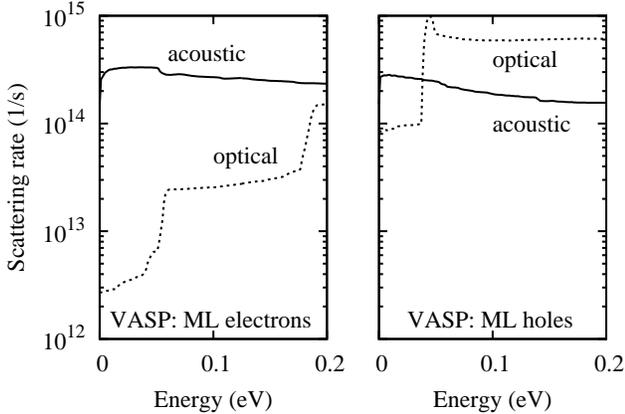}}
\caption{Electron-phonon (left) and hole-phonon (right) scattering rates in monolayer phosphorene at 300~K  where, the matrix element is obtained from VASP.}
\label{fig:P_mono_prates} 
\end{figure}

\begin{figure}[tb]
{\includegraphics{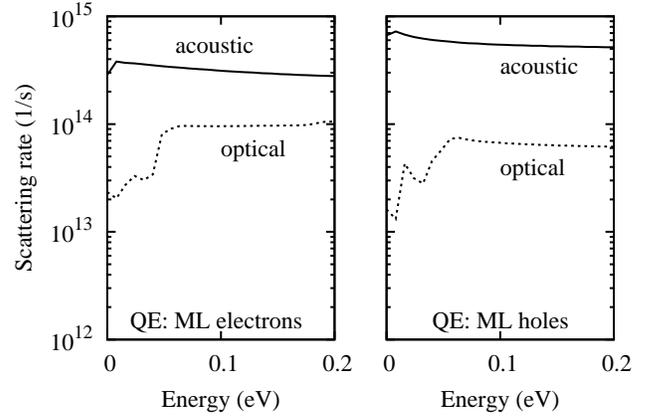}}
\caption{Electron-phonon (left) and hole-phonon (right) scattering rates in monolayer phosphorene at 300~K. The matrix elements have been calculated using QE.}
\label{fig:P_mono_prates_qe} 
\end{figure}

\begin{figure*}
    \centering
    \subfloat{\includegraphics{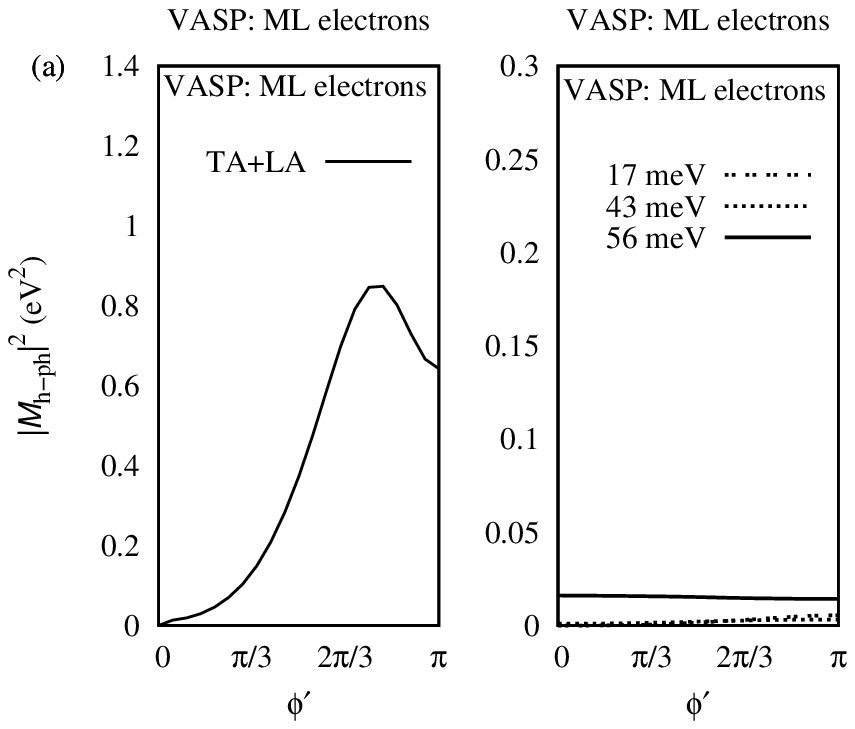}}
   \hfill
    \subfloat{\includegraphics{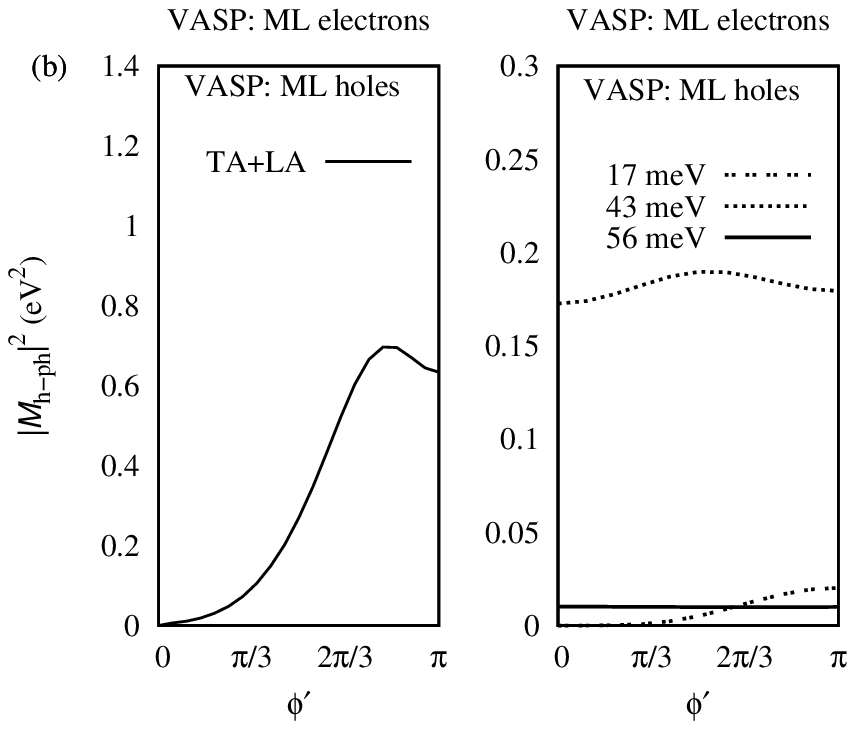}}
    \vfill
    \subfloat{\includegraphics{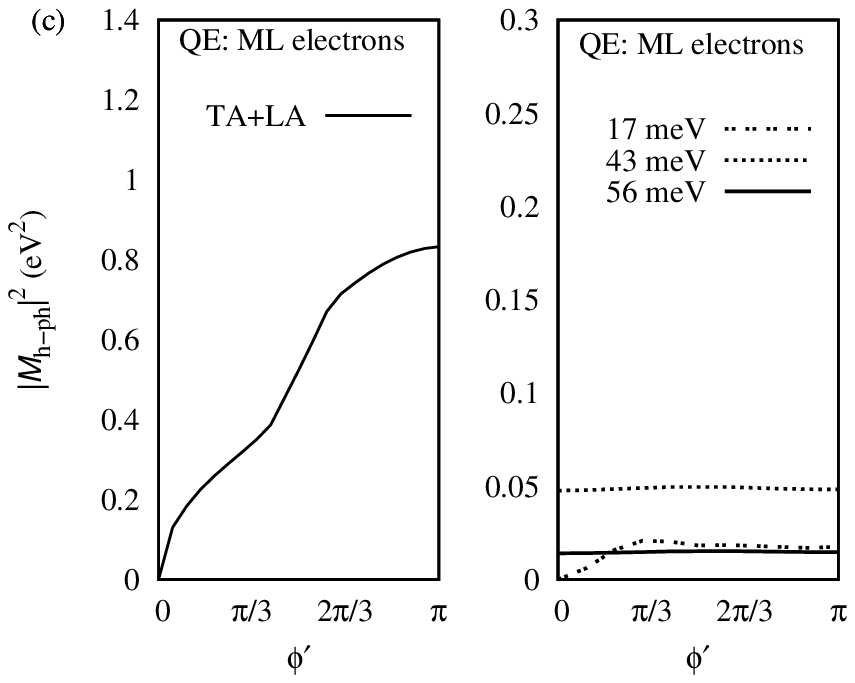}}
    \hfill
    \subfloat{\includegraphics{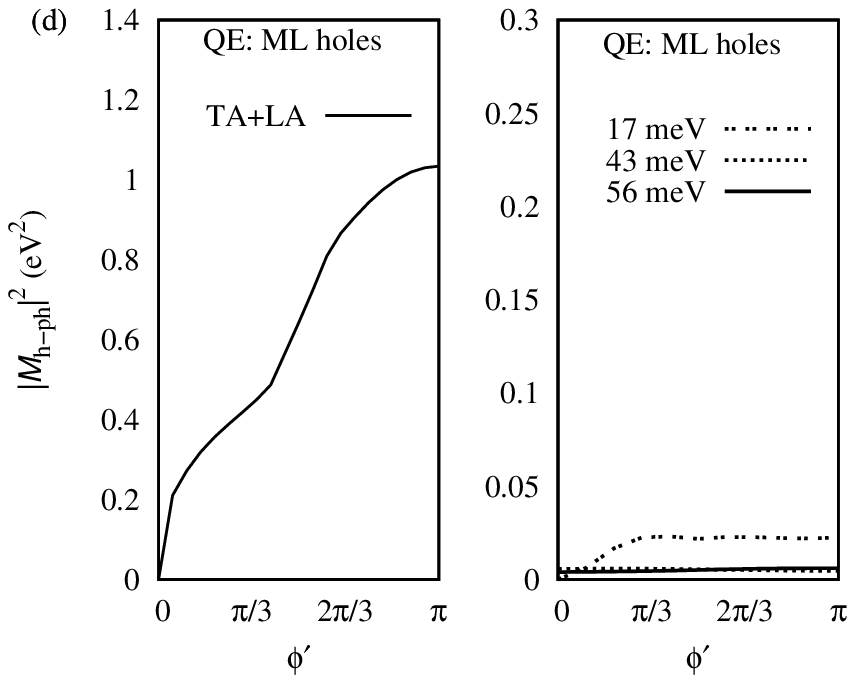}}
   \caption{(a) Acoustic (left) and optical (right) electron/hole-phonon matrix element
for electrons obtained from VASP,  (b) for holes obtained from VASP, (c) for electrons obtained from QE and (d) for holes obtained from QE, in monolayer phosphorene. The quantities
plotted here have been calculated following Eq.~(\ref{eq:Defpot}) for an initial and final electron
kinetic energies ($\approx{40~meV}$) as functions of the angle $\phi^{\prime}$
of the final wave vector ${\bf k}^{\prime}$ with the respect to the
armchair direction. The initial electron wave vector ${\bf k}$ is
taken to be along the armchair direction.}
    \label{fig:P_mono_DK}
\end{figure*}

\begin{figure*}
    \centering
        \subfloat{\includegraphics{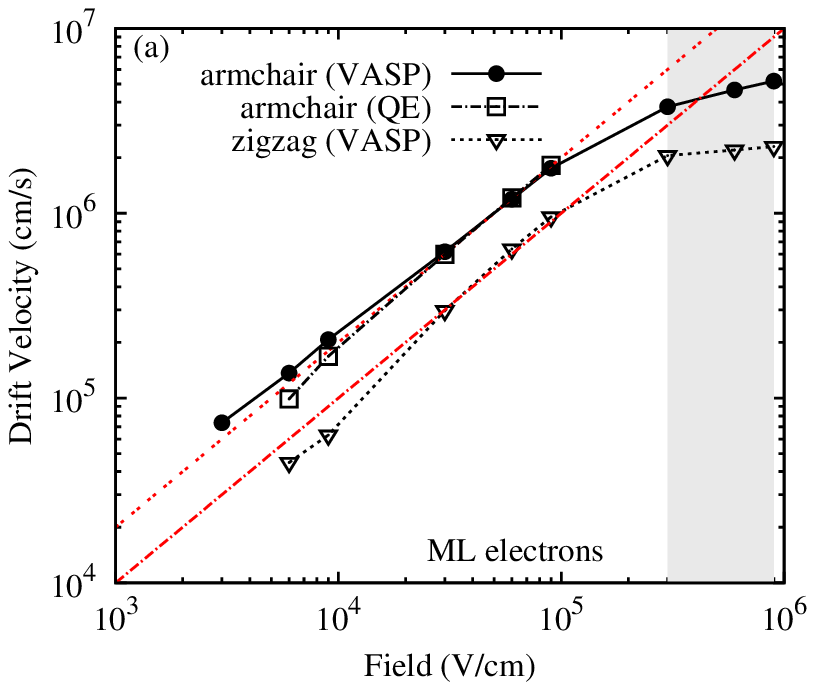}}
        \hfill
        \subfloat{\includegraphics{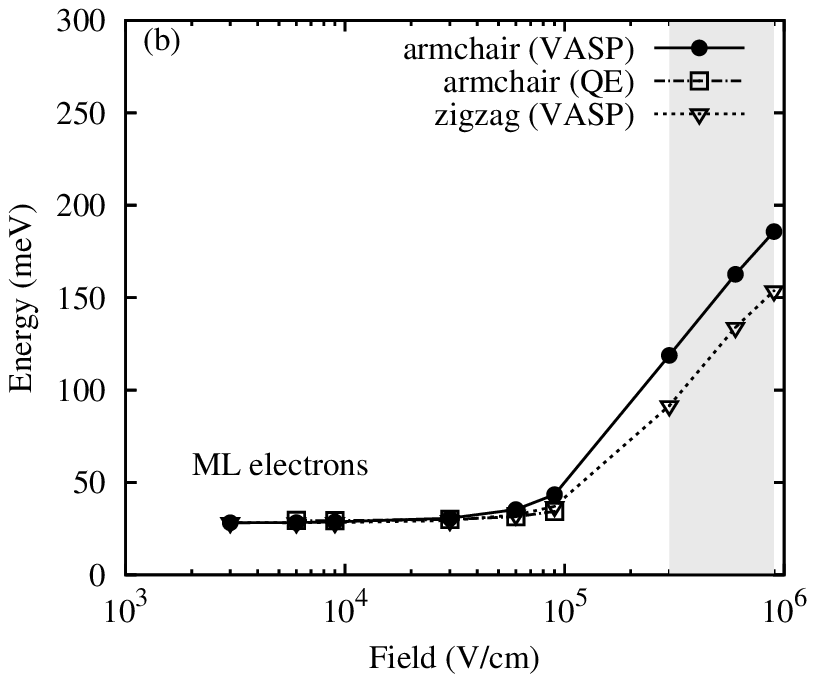}}
        \vfill
         \subfloat{\includegraphics{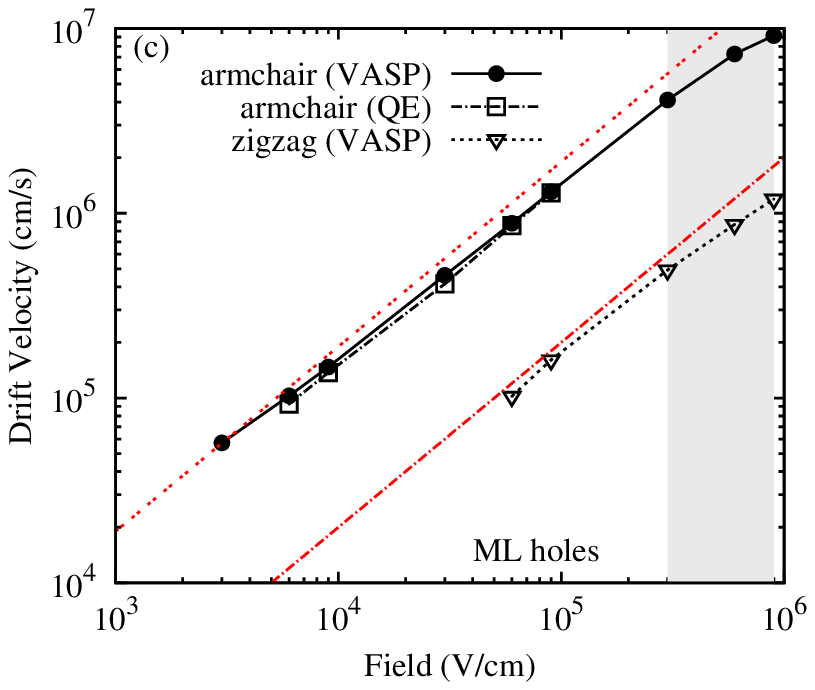}}
        \hfill
        \subfloat{\includegraphics{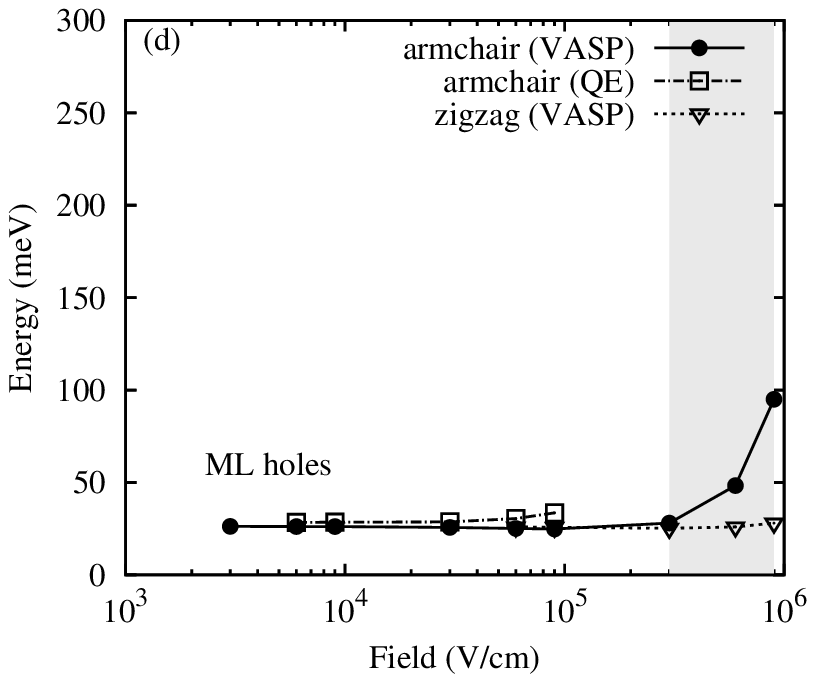}}
         \caption{(a) Drift-velocity \textit{vs.} field (left) and (b) average-energy \textit{vs.} field (right) characteristics
 at 300~K for electrons, and (c) and (d) for holes, in monolayer phosphorene calculated using full-band Monte Carlo simulations.
The dashed lines in the velocity plots show the Ohmic behavior based on the mobility determined from the zero-field
diffusion constant calculations. The electric
field is assumed to be along the armchair or zigzag direction, as
indicated. In the armchair direction, the electron-phonon matrix elements required for transport, are obtained using both VASP AND QE.
The shaded region in the high field regime indicates the approximate onset of scattering to higher bands, which was excluded from our model, results in this region only show a qualitative trend.}
   \label{fig:P_mono_highfield} 
\end{figure*}

In Figs.~\ref{fig:P_mono_prates} and \ref{fig:P_mono_prates_qe}, we show the angle-averaged 
scattering rates as a function of carrier kinetic energy obtained from VASP and Quantum Espresso,
both for electron-phonon and hole-phonon processes. Since phosphorene is a $\sigma_{{\rm h}}$-symmetric crystal,
the ZA phonons contribution is negligible to at first order in the electron-phonon interaction\cite{{Fischetti_2016}}.
Therefore electron/ZA-phonon and hole/ZA-phonon scattering has been ignored in our transport studies.
Figure~\ref{fig:P_mono_DK} shows the electron-phonon and hole-phonon matrix element, as in 
Eq.~(\ref{eq:Defpot}), calculated for an initial kinetic energy of $\approx$ 40~meV,
as a function of the final scattering angle $\phi^{\prime}$ 
(with respect to the armchair $\Gamma-X$ direction) for an initial wave vector
${\bf k}$ along the armchair direction. The highly anisotropic nature
of the electron/acoustic-phonon matrix elements indicates the importance
of using angular dependent deformation potentials, as already observed
in Ref.~\onlinecite{Liao_2015}.  For electrons, intravalley scattering
is dominated by in-plane acoustic phonons with strong backward scattering 
(Figs.~\ref{fig:P_mono_DK}a and \ref{fig:P_mono_DK}c). 
Intervalley scattering is controlled mainly by an optical phonon ($\approx$ 32~meV) 
with an intervalley deformation potential of about $1.7\times10^{9}$~eV/cm. 
For holes, when calculated from VASP, scattering is dominated by the intravalley ZO
mode with an energy of about 43~meV and a high deformation potential
of $1.7\times10^{9}$~eV/cm (Fig.~\ref{fig:P_mono_DK}b). However, when calculated from
Quantum Espresso, we notice that the in-plane acoustic modes dominate the scattering for holes((Fig.~\ref{fig:P_mono_DK}d)). 
Comparing these results to reports in literature using similar physical models, 
we note that Jin \textit{et al.}\cite{Jin_2016}, using Quantum Espresso, have concluded that 
acoustic phonons are the limiting factor in the hole transport, which is consistent with our 
Quantum Espresso results.
Note that this difference between VASP and QE does not translate to large differences in transport 
characteristics at room temperature since the 43~meV ZO mode has a low thermal population.

In Table~\ref{tab:mu_theory}, we list the values for the electron
and hole mobility we have obtained from the diffusion constant, both
in the armchair and the zigzag directions. We obtained very similar mobilities for
electrons and holes from both VASP and Quantum Espresso. We have already observed
that our results are significantly different from those presented
in Refs.~\onlinecite{Jin_2016} and \onlinecite{Liao_2015} and we
have speculated about possible causes for this difference. The velocity-field
and energy-field characteristics for electron and hole transport are shown in 
Fig.~\ref{fig:P_mono_highfield} for a uniform electric field applied along the armchair and zigzag
directions. For transport along the armchair direction, the low-field
mobility obtained for electrons from the velocity-field characteristics (Fig.~\ref{fig:P_mono_highfield}a) is in agreement
with both the Kubo-Greenwood results presented above as well as with
the value obtained from the diffusion constant. In contrast, for
electron transport along the zigzag direction, the Monte Carlo simulations
predict a low-field mobility a factor of $\approx$ 2 higher than
the analytic Kubo-Greenwood estimate. 

\begin{figure}[tb]
\vspace*{-1em}
\includegraphics{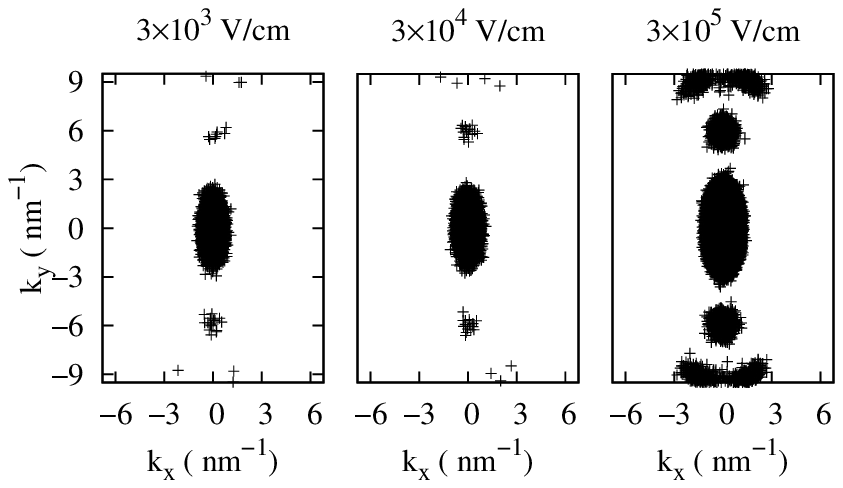} \caption{Distribution in reciprocal space for electrons in monolayer phosphorene for a field of $3\times10^{3}$
(left), $3\times10^{4}$ (center), and $3\times10^{5}$ V/cm (right),
along the armchair direction. Note the $\Gamma-Q$ intervalley transfer
at center, the $\Gamma-Y$, and $Q-Y$ intervalley transfer at the
highest field.}
\label{fig:P_mono_kspace} 
\end{figure}

We note a rather disappointing saturated velocity of $5\times10^{6}$~cm/s for electrons, especially for
a field along the zigzag direction. Interestingly, Fig.~\ref{fig:P_mono_kspace}
shows the occurrence of significant intervalley transfer to the $Q$
valley and even to the $Y$ valley. However, this does not translate
into any negative differential mobility, since the effective masses
in these satellite valleys are similar to those in the $\Gamma$ valley. 
The hole mobility obtained from the velocity-field characteristics (Fig.~\ref{fig:P_mono_highfield}c) and 
diffusion constant in the armchair direction is about the same as for electrons. 
However, the hole mobility is significantly lower along the zigzag direction due to 
`flatness' of the valence band along the zigzag direction. 
The velocity tends to deviate from linearity for electrons at very high fields. However,
we do not observe any saturation even at high fields of ~$10^{5}$~V/cm. 
At low fields, the velocities for both electrons 
and holes in the zigzag direction are lower than the thermal
velocity, making it difficult to extract the mobility from the velocity-field characteristics 
numerically. Up to a field of 100 kV/cm, the average carrier energy for electrons remains at the thermal energy ($\approx$ 25~meV) and the electrons are in the ohmic regime (Fig.~\ref{fig:P_mono_highfield}b). For holes, the ohmic regime extends to even higher fields, especially in the zigzag direction (Fig.~\ref{fig:P_mono_highfield}d).

\subsection{Phosphorene bilayers}

\label{sec:P-bi} 
\begin{figure}[tb]
\hbox{\includegraphics{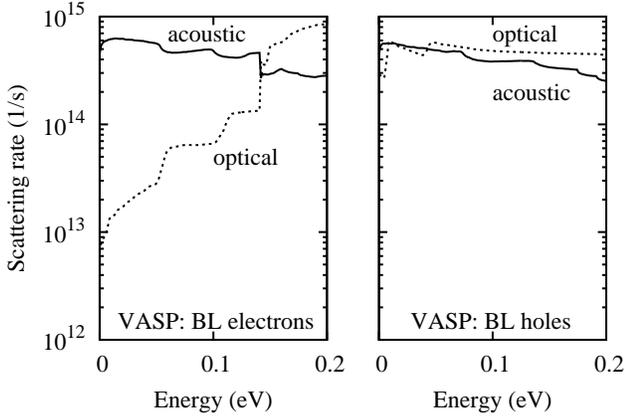}}
\caption{Electron-phonon (left) and hole-phonon (right) scattering rates (VASP) in bilayer phosphorene at 300~K.}
\label{fig:P_bi_prates} 
\end{figure}

\begin{figure*}
    \centering
    \subfloat{\includegraphics{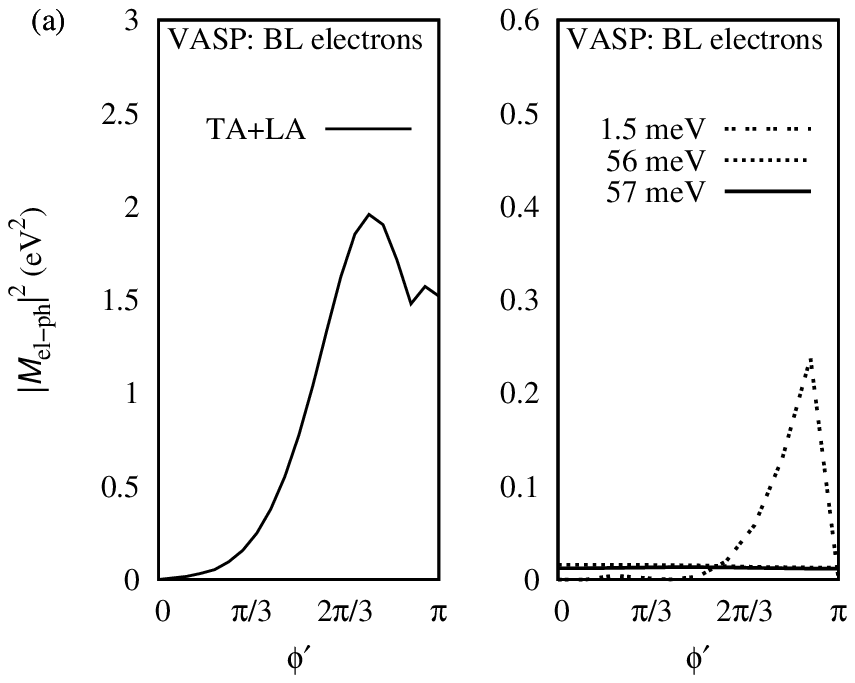}}
  \hfill
    \subfloat{\includegraphics{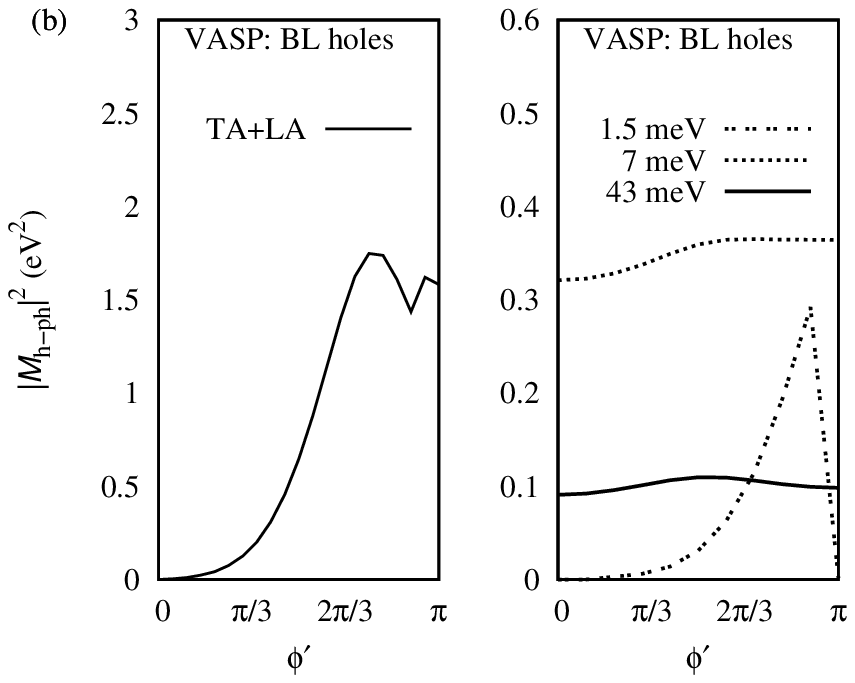}}
   \caption{(a) Acoustic (left) and optical (right) electron/hole-phonon matrix element
for electrons and, (b) for holes, in bilayer phosphorene (VASP).}
    \label{fig:P_bi_DK}
\end{figure*}

\begin{figure*}
    \centering
   
     \subfloat{\includegraphics{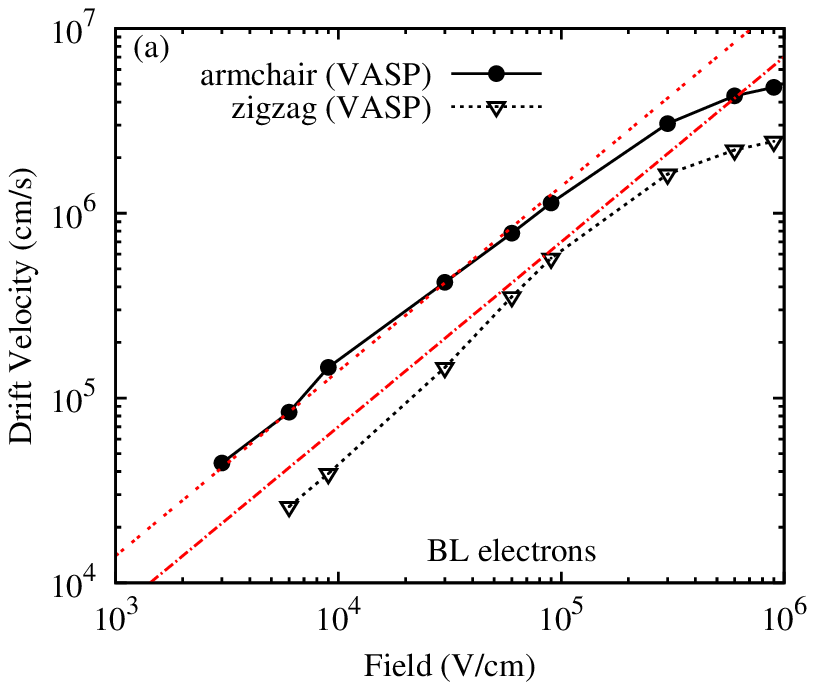}}
     \hfill
     \subfloat{\includegraphics{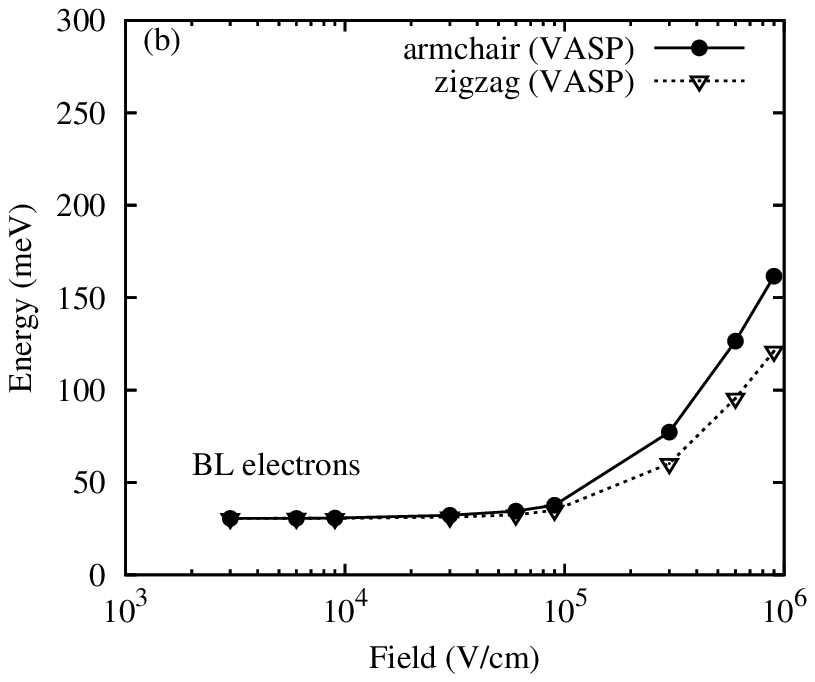}}
     \vfill
     \subfloat{\includegraphics{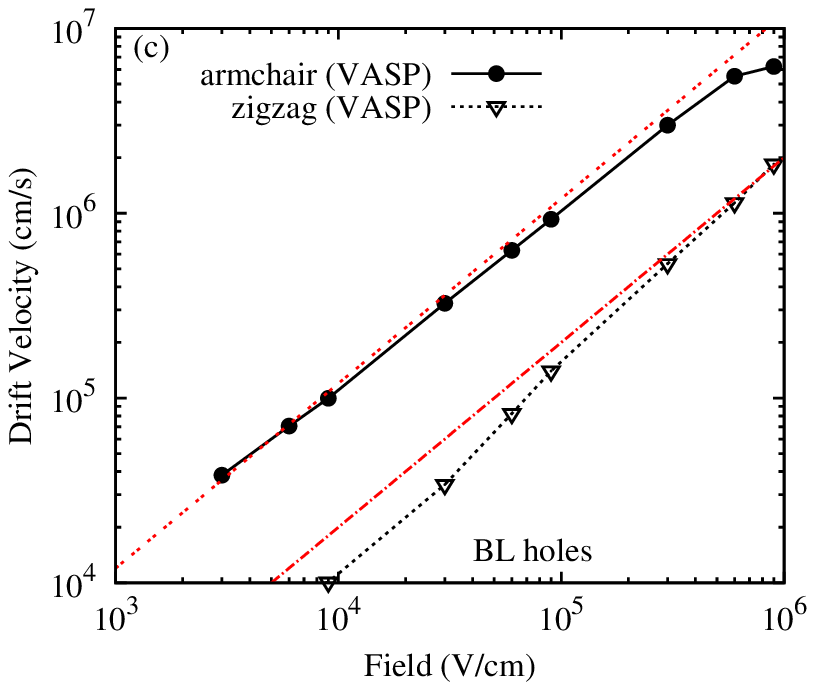}}
     \hfill
     \subfloat{\includegraphics{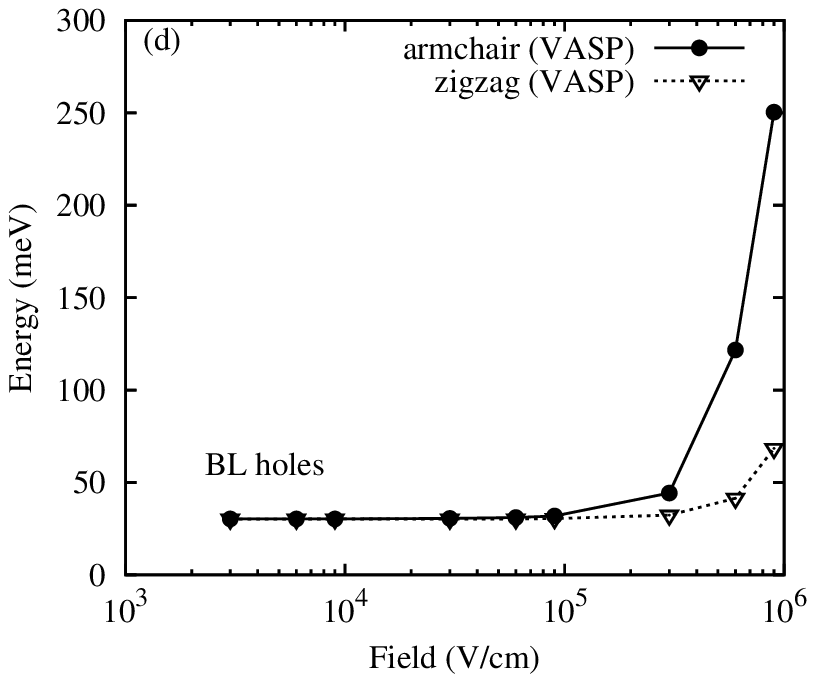}}
    \caption{(a) Drift-velocity \textit{vs.} field (left) and (b) average-energy \textit{vs.} field (right) characteristics
 at 300~K for electrons and, (c) and (d) for holes, in bilayer phosphorene (VASP) calculated using full-band Monte Carlo simulations.}
    \label{fig:P_bi_highfield}
\end{figure*}
For bilayers, the angle-averaged scattering rates as a function of carrier kinetic
energy are shown in Fig.~\ref{fig:P_bi_prates}. Similar to monolayers, scattering with ZA phonons is ignored.
Fig.~\ref{fig:P_bi_DK} shows the electron/hole-phonon matrix element, 
Eq.~(\ref{eq:Defpot}), at a small electron kinetic energy, plotted as
a function of final scattering angle $\phi^{\prime}$ for an initial
wave vector ${\bf k}$ along the armchair direction.
Similar to our findings for electrons in monolayers, intravalley
scattering is controlled to a large extent by in-plane acoustic phonons 
but with a stronger backward scattering.
However, the many optical phonons seen in Fig.~\ref{fig:P_bi_DK}a (right frame)
all contribute significantly to intravalley processes. For intervalley
scattering, similar to monolayer, the optical mode with an energy of 32~meV exhibits the largest 
deformation potential, $\approx 1.7\times10^{9}$~eV/cm.
However, among the optical modes, because of their low energies, scattering is 
dominated by a low-energy inter-layer optical mode ($\approx$ 1.5~meV), 
despite its low deformation potential (ranging from $5.4\times10^{3}$ to $2\times10^{8}$~eV/cm). 
This is due to the large phonon population and the large amplitude of the squared matrix element that is inversely proportional to the phonon energy.
For holes, as seen in Fig.~\ref{fig:P_bi_DK}b (left frame), intravalley scattering is 
dominated by strong backward scattering due to acoustic modes and low-energy inter-layer 
optical modes with an energy of about 1.5~meV and 7~meV. The associated the deformation potentials 
reach maximum values of about $2.2\times10^{8}$~eV/cm and $6.2\times10^{8}$~eV/cm, respectively. 
 
In Fig.~\ref{fig:El_ph_bands}d, we show that the acoustic modes have multiple branch crossings 
with the low-energy optical modes, making it difficult or impossible to distinguish 
between acoustic and optical modes. Therefore, we believe that the contribution by low-energy 
optical modes might be much higher than what we observe in Fig.~\ref{fig:P_bi_DK}.
Note that our conclusion that low-energy inter-layer optical modes play such
a dominant role in bilayers is not consistent with the results presented
by Jin \textit{et al.}\cite{Jin_2016}: Their results show that acoustic phonons
dominate the scattering for both electrons and holes and that 
total scattering rates are much lower than we obtained, leading to higher mobility 
for bilayer phosphorene.
However, as we have mentioned above, a strong coupling
between electron and inter-layer optical phonons has been previously
shown in the case of graphene bilayers\cite{Park_2008}.\\

\begin{figure}[tb]
\vspace*{-1.0em}
\includegraphics{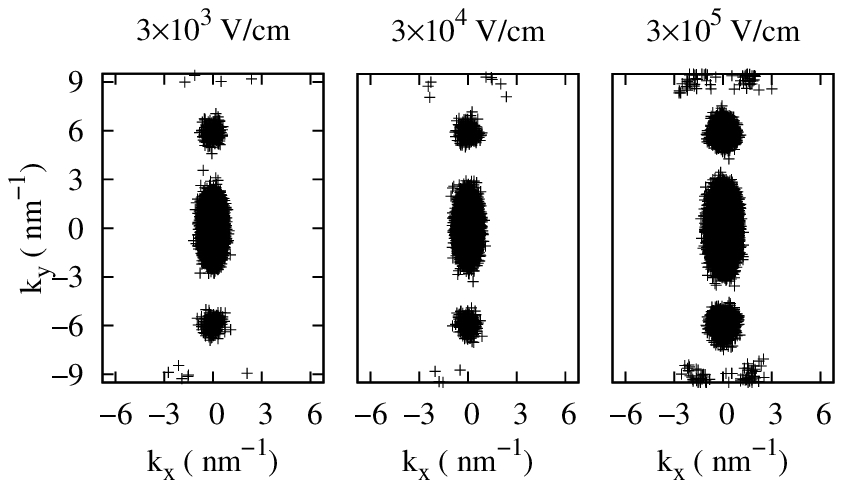}
\caption{Distribution in reciprocal space of electrons in bilayer phosphorene
at high electric fields along the armchair direction at 300~K.}
\label{fig:P_bi_kspace} 
\end{figure}

Finally, the velocity-field and energy-field characteristics for electron and hole transport in 
bilayers are shown in Fig.~\ref{fig:P_bi_highfield}. 
The carrier mobilities for electrons and holes obtained from velocity-field characteristics (Fig.~\ref{fig:P_bi_highfield}a and Fig.~\ref{fig:P_bi_highfield}c) 
along the armchair and zigzag directions are in agreement with the values obtained from the 
diffusion constants (Table~\ref{tab:mu_theory}).
As seen in monolayers, the saturated velocity for electrons in bilayers is 
relatively low as well, which can be attributed to intervalley transfer to 
$Q$ and $Y$ valleys (Fig.~\ref{fig:P_bi_kspace}). The hole mobility in the armchair
direction is about the same as for electrons, whereas in the zigzag direction, 
the mobility is significantly lower due to `flatness' of the valence band along 
the zigzag direction. The results we have obtained show that the carrier mobility 
decreases slightly when moving from monolayers to bilayers. This can be attributed to 
scattering by low energy inter-layer optical modes. 
We believe that in the presence of a substrate and/or a gate dielectric, 
the mobility in bilayers should increase, especially for holes, thanks to the dampling -- or even suppression -- of 
scattering with low-energy inter-layer optical modes.
Similar to monolayers, the ohmic regime, extends up to a field of 100 kV/cm for electrons (Fig.~\ref{fig:P_bi_highfield}b)
 and it extends to higher fields for holes (Fig.~\ref{fig:P_bi_highfield}d).

We should note again that our results are in disagreement with
those reported by Jin \textit{et al.}\cite{Jin_2016}, despite the
similarity of the methods and assumptions made. This is the case for
both monolayers and bilayers. We only note that we have observed a
notable dependence of the computed mobility on the discretization
employed, as mentioned previously, mandating a very fine mesh in reciprocal
space. A different band structure, resulting from different `flavors'
of the DFT numerical method used, may also be a plausible cause for
this difference that, ultimately, we cannot explain.

\subsection{Thickness dependence of transport properties}
\label{sec:P-thickness} 

In light of the observation we have made in the previous section regarding
the low-field carrier mobility in mono- and bi-layers, it is interesting
to discuss more generally the dependence of the carrier mobility on
the thickness of phosphorene multi-layers and bulk black phosphorus (bP).

As we have mentioned in Sec.~\ref{sec:Review}, experiments show
a hole mobility that is strongly dependent on thickness\cite{Liu_2014,Xia_2014},
with values hovering around several hundreds cm$^{2}$V$^{-1}$s$^{-1}$
in thick layers, sharply decreasing in layers thinner than 10~nm,
being as low as 1-10 cm$^{2}$V$^{-1}$s$^{-1}$ in layers 2-3~nm-thin\cite{Li_2014}.
Such a behavior can be understood as the result of several effects:
An increasing effective mass, a stronger carrier confinement, a larger
deformation potential, and softening of the inter-layer optical
phonons as we move from bP to monolayers. We discuss each of these
effects in turn. 

\textit{1. Thickness dependence of the carrier effective mass.} We
have calibrated the local empirical pseudopotentials for P given in
Ref.~\onlinecite{Bellaiche_1996} to reproduce the band structure
of monolayer phosphorene 
\begin{equation}
V_{{\rm P}}(q)\ =\ \sum_{j=1}^{4}\ a_{j}e^{-b_{j}(q-c_{j})^{2}}[1-d_{j}e^{-f_{j}q^{2}}]\ ,\label{eq:pseudo_P}
\end{equation}
with parameters $a_{j}$, $b_{j}$, $c_{j}$, $d_{j}$, and $f_{j}$
given in Ref.~\onlinecite{Bellaiche_1996} except for $b_{1}=0.834517$
and $a_{4}=0.085232$ (in atomic and Rydberg units). The calibration
has been performed using an energy cutoff of 5~Ry. The modifications
of $b_{1}$ and $a_{4}$ have been made to obtain the desired band
gap for monolayer phosphorene\cite{note_1} ($\approx1.5$ eV from
Refs.~\onlinecite{Qiao_2014,Tran_2014,Zhang_2014}). The electron and
hole effective masses we have obtained for bulk black phosphorus and
of 1-, 2-, and 3-layer phosphorene are shown in Tables~\ref{tab:emass}
and \ref{tab:hmass}, respectively. We have already observed that obtaining a meaningful
effective mass for holes along the zigzag direction is not possible,
since the dispersion of the valence band along this direction ($\Gamma-Y$
in few-layer phosphorene, $Z-A$ in bulk bP) is very flat and extremely
sensitive on the way the calculations are performed, as also remarked
in Ref.~\onlinecite{Lew_2015}. Indeed, in some cases, the valence-band
(VB) maximum is found slightly away from the $\Gamma$ point~\cite{Lew_2015,Rodin_2014,Li_2014a,Peng_2014},
thus rendering the band gap of the material slightly indirect. Therefore,
extracting a `curvature' effective mass is impossible. In order to
bypass this problem, we have calculated a conductivity mass (\textit{i.e.},
the slope of the dispersion) along that direction over an energy window
of $k_{{\rm B}}T\approx$ 25 meV as $m_{{\rm h}}=\hbar k_{{\rm kT}}/\sqrt{50\,{\rm meV}}$
where $E(k_{{\rm kT}})=25\,{\rm meV}$.  The thus obtained conductivity
mass will give an idea about the group velocity that enters the Kubo-Greenwood
formula for the mobility. (Obviously, the high density-of-states associated
with the flat dispersion still strongly depresses the mobility via
the large momentum-relaxation rates).

The results are listed in Table~\ref{tab:hmass}.
Tables~\ref{tab:emass} and \ref{tab:hmass} also list the values
of the effective masses we have obtained using DFT (VASP) calculations.
These values are affected by a larger discretization error and are
expected to be somewhat smaller compared to experimental results because
of the smaller band gap predicted by DFT calculations. We see that
the effective mass decreases with increasing thickness for both electrons
and holes. This leads us to expect an increasing mobility when moving
from monolayers to bulk black phosphorus. However, the masses decrease
by an amount that is too small to explain the difference between bilayer
and monolayer mobility behavior seen in Table~\ref{tab:mu_exp}.
\begin{table}
\caption{Electron effective mass (in units of the free electron mass) in phosphorene
layers and bulk black phosphorus calculated using empirical pseudopotentials
and DFT (in parentheses).}
\begin{ruledtabular}
\begin{tabular}{ccccccccc}
 & $\Gamma-X$  & $\Gamma-Y$  & $\Gamma-S$  & DoS  &  &  &  & \tabularnewline
\hline 
monolayer  & 0.22 (0.14)  & 1.1 (1.24)  & 0.46  & 0.48  &  &  &  & \tabularnewline
bilayer  & 0.21 (0.10)  & 1.1 (1.33)  & 0.44  & 0.47  &  &  &  & \tabularnewline
trilayer  & 0.21  & 1.1  & 0.44  & 0.47  &  &  &  & \tabularnewline
\hline 
 & $Z-Q$  & $Z-A$  & $Z-\Gamma$  & DoS  &  &  &  & \tabularnewline
\hline 
bulk  & 0.09  & 1.1  & 0.16  & 0.25  &  &  &  & \tabularnewline
bulk (exp)$^{(a)}$  & 0.0826  & 1.027  & 0.128  &  &  &  &  & \tabularnewline
\end{tabular}
\end{ruledtabular}

\begin{raggedright}
(a) Experimental data from Ref.~\onlinecite{Morita_1986} 
\par\end{raggedright}
\label{tab:emass} 
\end{table}

\begin{table}
\caption{Hole effective mass (in units of the free electron mass)
in phosphorene layers and bulk black phosphorus calculated using empirical
pseudopotentials and DFT (in parentheses).}
\begin{ruledtabular}
\begin{tabular}{ccccccccc}
 & $\Gamma-X$  & $\Gamma-Y^{{\rm (a)}}$  & $\Gamma-S$  & DoS  &  &  &  & \tabularnewline
\hline 
monolayer  & 0.20 (0.14)  & 1.7  & 0.59  & 0.59  &  &  &  & \tabularnewline
bilayer  & 0.19 (0.09)  & 1.2 (3.08)  & 0.56  & 0.51  &  &  &  & \tabularnewline
trilayer  & 0.21  & 0.91  & 0.62  & 0.49  &  &  &  & \tabularnewline
\hline 
 & $Z-Q$  & $Z-A$  & $Z-\Gamma$  & DoS  &  &  &  & \tabularnewline
\hline 
bulk  & 0.09  & 0.51  & 0.37  & 0.26  &  &  &  & \tabularnewline
bulk (exp)$^{(b)}$  & 0.076  & 0.648  & 0.280  &  &  &  &  & \tabularnewline
\end{tabular}
\end{ruledtabular}

\begin{raggedright}
(a) Optical (slope) mass at 25 meV. The curvature mass along the $\Gamma-Y$
direction cannot be defined at $\Gamma$ due to the `flatness' of
the dispersion.\\
 (b) Experimental data from Ref.~\onlinecite{Morita_1986} 
\par\end{raggedright}
\label{tab:hmass} 
\end{table}

\begin{figure}[tb]
\includegraphics[width=8.5cm]{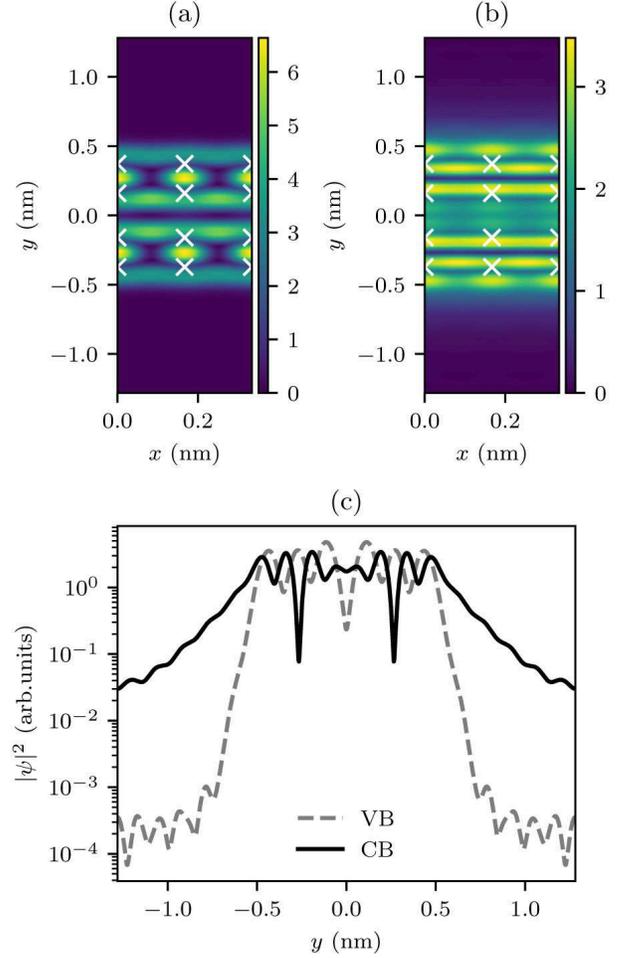}
 \caption{(a) Squared amplitude of the wavefunction for the highest-energy valence
band and (b) for the lowest-energy conduction band, averaged over
a unit cell along the armchair direction for a phosphorene
bilayer. A vacuum of 2.6~nm was used in the construction of the supercell. (c) The same quantity, but averaged also along the zigzag
direction. The cross denote the position of the phosphorus atoms.
These results have been obtained using local empirical pseudopotentials.
Note the large penetration of the conduction-band wavefunction into
the interlayer region.\protect \\
  }
\label{fig:PBL_wf} 
\end{figure}

\begin{figure}[tb]
\includegraphics[width=8.5cm]{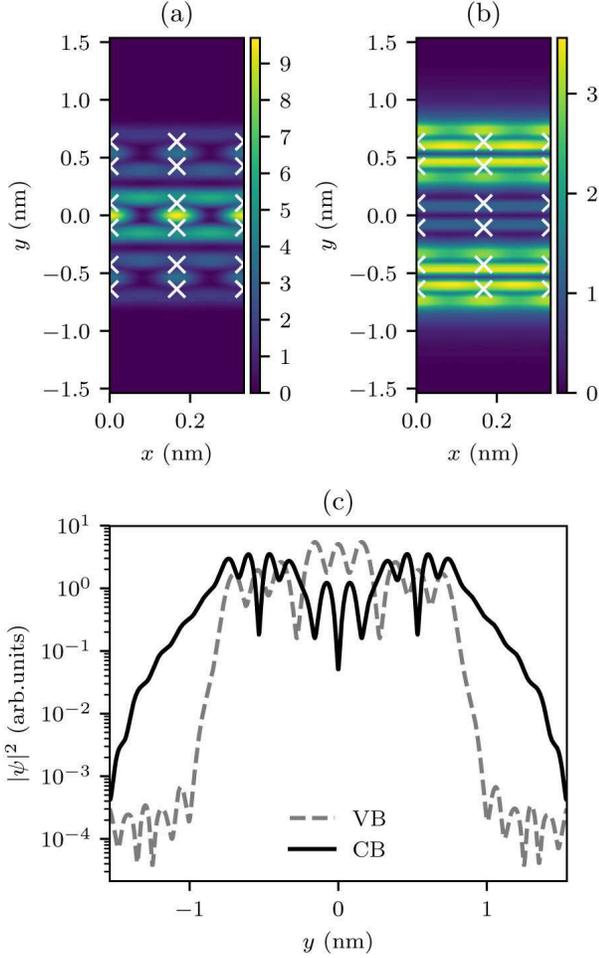}
\caption{As in Fig.~\ref{fig:PBL_wf}, but for a phosphorene trilayer. 
A vacuum of 3.1~nm was used in the construction of the supercell. Note
how the valence-band wavefunction resembles a cosine-like standing
wave, with a maximum in the central region, whereas the conduction-band
wavefunctions shows two broad peaks, as for the first excited state
of a particle in a box. Therefore, the envelopes of these wavefunctions
appear to be set more by the top and bottom trilayer surfaces than
by the ionic potentials in each layer.\protect \\
  }
\label{fig:PTL_wf} 
\end{figure}

\begin{figure}[tb]
\includegraphics[width=8.5cm]{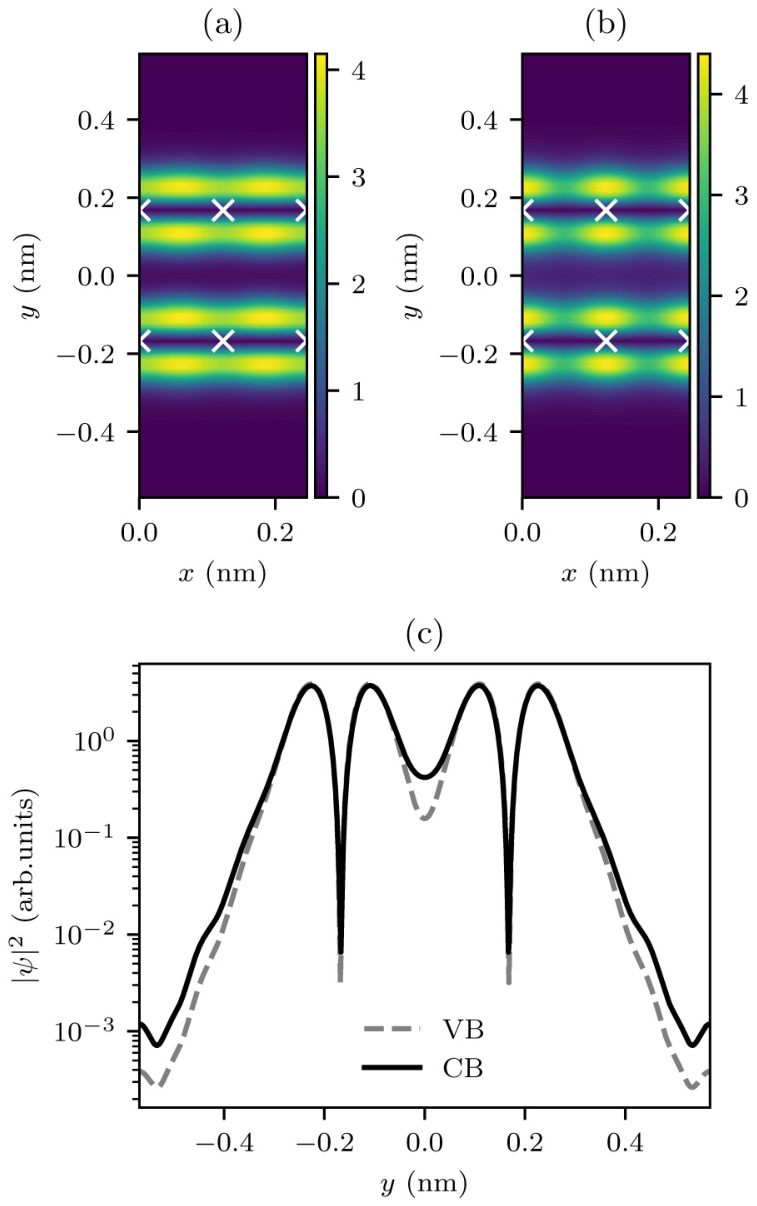} 
\caption{As in Fig.~\ref{fig:PBL_wf}, but for a graphene
bilayer. A vacuum of 1.2~nm was used in the construction of the supercell. Note how also conduction-band wavefunction vanishes in the
interlayer region, in striking contrast with the behavior observed
for bilayer phosphorene, Fig.~\ref{fig:PBL_wf} (b). The envelopes
of these wavefunctions appear to be confined in the out-of-plane direction
exclusively by the ionic potentials in each layer.}
\label{fig:GBL_wf} 
\end{figure}

\textit{2. Quantum confinement effects.} Intrinsically 2D materials have, among
their many remarkable properties, two major advantages for the purpose
of electronic devices when compared to thin `3D materials'. In thin
3D materials, quantum confinement is the result of conduction-band
discontinuities with insulators (as for thin Si films confined by
SiO$_{2}$ or other oxides) whereas in 2D materials, electrons are
naturally confined in two dimensions by ionic potentials (as for graphene).
Bulk 3D material mobilities range from respectable to excellent,
but in thin films the carrier mobility is severely depressed by `wavefunction-overlap'
effects. On the other hand, 2D materials have the potential of exhibiting
excellent mobilities even in the limit of atomic thickness.  Moreover,
confinement by insulators leads to a problematic energetic up-shift
of the ground-state subbands in semiconductor ultra-thin films or
nanowires based on 3D materials\cite{Fischetti_2013a}, causing a
potentially unacceptably high gate-current leakage. Again, the naturally
ionic confinement in 2D materials will save them from such an unacceptable
high gate leakage current.

Two-dimensional $sp^{2}$ (or $sp^{2}/sp^{3}$) group-IV materials
and transition-metal dichalcogenides (TMDs) are typical examples of
materials that exhibit these desired properties. In these materials,
the presence of a lone electron (such as the electron occupying the $p_{z}$ orbital in $sp^{2}$
group-IV materials) or the absence of unpaired electrons (such as
in TMDs, with the 2-4 $p$ electrons from the transition metal and the
6 hybrid-$sp$ electrons from the two chalcogens) results in either
a $\pi$-band strongly localized in the out-of-plane direction (out
of the ``lone'' $p_{z}$ electron in the $sp^{2}$ group-IV 2D crystals),
or in out-of-plane-localized states inside the 2D crystal (TMDs).
In a multi-layer van der Waals heterostructure, `vertical' band-discontinuities
at the top and bottom surfaces or interfaces, do not significantly
affect the band structure. Indeed, the van der Waals coupling between
adjacent layers is too small (of the order of a few tens of meV per
atom both in graphene\cite{Charlier_1994} and TMDs\cite{Bjorkman_2012})
to alter the bonding properties and, most important, the out-of-plane
localization of the wavefunctions. These materials are indeed easily
exfoliated and are relatively chemically stable \textendash{} a result
of the out-of-plane localization of the valence-band states. Moreover,
they show a very low out-of-plane conductivity \textendash{} a result
of the out-of-plane localization of the conduction-band states. However,
the in-plane mobility remains high, even in monolayers. Indeed, graphene
exhibits an extremely high mobility, $\approx10^{5}$ cm$^{2}$V$^{-1}$s$^{-1}$.
Such a high mobility, even for carriers confined over a thickness
of the order of 0.1~nm, is the result not only of pseudospin-conservation
and reduced back-scattering, but also of the fact that the scattering
form-factor of the wavefunctions does not depend on the confinement
along the out-of-plane direction.

The recent studies by Qiao \textit{et al.}\cite{Qiao_2014} and by
Xin-Hu and co-workers\cite{Xin-Hu_2016} have shown that phosphorene
(and, we speculate, other group-V materials, such as also arsenene
and 
antimonene) is remarkably different: In multilayers, the lone $p$-orbital
pairs hybridize into bonding and antibonding orbitals, weakly localized
or even delocalized along the out-of-plane direction, with wavefunctions
that spread across the interlayer gap. The interlayer coupling energy
is now much larger: A value of 0.46 eV has been used for the interlayer
coupling energy in calibrated tight-binding calculations of phosphorene\cite{Fukuoka_2015},
rendering this a `quasi-covalent' bonding. This should be compared
to the pure van der Waals interaction energy, of the order of a tens
of meV/atom, as mentioned above. Such an effect, recently discussed
at length by Xin-Hu \textit{et al.}\cite{Xin-Hu_2016}, has been shown
to be responsible for the poorly-understood behavior of phonon frequencies
in few-layer phosphorene. These considerations imply that now the
hybridized valence- and conduction-band wavefunctions can spread significantly
across the interlayer. An additional strong hint that phosphorene
multilayers are quite different from pure van der Waals materials
is provided by the fact that that in bulk bP, the electron mobility
along the $z$-axis (perpendicular to the plane of the layers) is
about the same as along the zigzag direction: 400 cm$^{2}$V$^{-1}$s$^{-1}$
(out-of-plane) \textit{vs.} 460 cm$^{2}$V$^{-1}$s$^{-1}$ (zigzag)
at 200~K, compared to 2,300 cm$^{2}$V$^{-1}$s$^{-1}$ along the
armchair direction\cite{Akahama_1983}. The situation is similar for
holes, their mobility being 540 cm$^{2}$V$^{-1}$s$^{-1}$ along
the out-of-plane direction, 1,300 cm$^{2}$V$^{-1}$s$^{-1}$ along
the zigzag direction, and 3,300 cm$^{2}$V$^{-1}$s$^{-1}$ along
the armchair direction, also at 200~K (Ref.~\onlinecite{Akahama_1983}).
This means that carriers move from one layer to the next with relative
ease. On the contrary, for graphite, the measured in-plane electron
mobility is quite high, about 13,000 cm$^{2}$V$^{-1}$s$^{-1}$ at
300~K, but the material behaves essentially as an insulator along
the out-of-plane direction, with a conductivity (mobility) about 3,000
times lower\cite{Pierson_1993,Dresselhaus_2002}. This means that
carriers are not transferred between layers, instead being localized
within each layer.

To illustrate the non-van der Waals nature of phosphorene, in Figs.~\ref{fig:PBL_wf}
and \ref{fig:PTL_wf}, we show the squared amplitude of the wavefunctions
corresponding to the highest-energy valence-band state and lowest-energy
conduction-band state in bilayer and trilayer phosphorene. (The van
der Waals gap of 3.20~\AA~reported in Ref.~\onlinecite{Qiao_2014}
has been assumed in these empirical-pseudopotential calculations.)
The wavefunction can be seen to `spill' into the inter-layer region,
especially for the conduction-band in bilayers (Fig.~\ref{fig:PBL_wf}
(b) and (c)). Note also in Fig.~\ref{fig:PTL_wf} how the valence-band
wavefunction reaches a maximum amplitude in the central layer, as
for the ground-state of a particle-in-a-box textbook problem. The
next excited state, the wavefunction of the conduction band, exhibits
instead two `bumps', once more reminding us of the first excited state
of confined bulk-like system compared to a 2D van der Waals material.
As just mentioned, this indicates that the electronic states are more
sensitive to the boundary conditions at the surfaces of the top and
bottom layers than to the ionic potentials in each layer. 
To contrast how wavefunctions manifest themselves in a van der Waals
material, we show the valence- and conduction-band wavefunctions in
bilayer graphene (obtained using the empirical pseudopotentials given
in Ref.~\onlinecite{Fischetti_2013} with a van der Waals gap of 3.35~\AA)
in Fig.~\ref{fig:GBL_wf}. In this case the wavefunctions vanish
in the inter-layer region, showing that in these $sp^{2}$ layers
the ionic potentials strongly confine the 2DEG within each layer.

This discussion suggests that multi-layer phosphorene behaves more
like a bulk covalent material than a van der Waals system, as already
argued by Qiao \textit{et al.}\cite{Qiao_2014} and Xin-Hu and co-workers\cite{Xin-Hu_2016}:
The `vertical' confinement of the 2DEG appears to be mainly controlled
not by the ionic potentials but by the conduction/valence-band discontinuities
at top/bottom interfaces. This situation is similar to the case of
semiconductor thin films or quantum wells. In such cases, carrier-phonon
scattering is mainly 
controlled by the well-known scattering form factor (a functional
of the `envelope' wavefunction along the out-of-plane direction $z$,
$\zeta_{n}(z)$, for carriers in band $n$, averaged over an in-plane
unit cell), 
\begin{equation}
\mathcal{I}_{nm}\ =\ \int\ {\rm d}z\ |\zeta_{n}(z)|^{2}\ |\zeta_{m}(z)|^{2}\ ,\label{eq:form-fact}
\end{equation}
originally derived by Price\cite{Price_1981}. This form factor increases
as $1/W^{2}$ as the well-width $W$ (or multilayer thickness, in
our case) decreases, resulting in a carrier mobility that vanishes
as $W^{2}$ for small $W$. In a way, we are
subjected to the same problems that govern Si thin film mobility:
As the thickness decreases, the mobility is depressed because of this
undesired `form-factor' effect. In phosphorene, this may not be an
effect as strong in the case of fully delocalized states, as for Si
or Ge thin bodies, but it is also far removed from the more ideal
cases of graphene, other group-IV monolayers, or TMDs.

\begin{table}[]
\centering
\caption{Form-factor of the wavefunctions in mono-, bi-, tri-layer phopshorene for the valence band maximum (VBM) and conduction band minimum (CBM)}
\begin{ruledtabular}
\label{tab:formfactor}
\begin{tabular}{clclc}
$I_{mn}$ (m = n)& VBM  & CBM   \\
\hline
 monolayers& 2.2417 &  1.4081  \\
 bilayers& 1.1606 &  0.7992  \\
 trilayers& 0.9463 & 0.6752   \\
 
\end{tabular}
\end{ruledtabular}
\end{table}

Calculated form-factors for the wavefunctions in
1-, 2-, 3-layer phosphorene for valence band maximum (VBM) and 
conduction band minimum (CBM) are shown in Table~\ref{tab:formfactor}.
As expected, the form-factor decreases with increasing in thickness, however,
they decrease by not more than a factor of 2. Therefore, going from mono- to bi-layer phosphorene, we expect the mobility to increase by not more than a
factor of 4 (the areal mass density in bilayers is twice that of monolayers). On the contrary, Cao \textit{et al.}\cite{Cao_2015} observed an increase in mobility by a factor of 80
moving from 1- to 2-layer phosphorene.

\textit{3. Deformation potentials.} A survey of the literature shows
that the band structure changes with strain in a way that depends
on the thickness of the system. In monolayers, a 4-6\% biaxial strain
can have the huge effect of reversing the anisotropy\cite{Fei_2014},
whereas in bulk bP strain seems to have a weaker effect\cite{Fukuoka_2015}.
This seems to suggest that the deformation potentials decrease with
increasing thickness, an effect that may contribute to boosting the
carrier mobility in thicker films. 
We have confirmed this by using the empirical pseudopotential given
by Eq.~(\ref{eq:pseudo_P}) to calculate the change of the band gap,
$E_{{\rm G}}$, under hydrostatic stress, $V_{0}{\rm d}E_{{\rm G}}/{\rm d}V$
(where $V_{0}$ and $V$ are the volume of the relaxed and hydrostatically
strained crystal, respectively) for monolayer phosphorene and bP.
This quantity yields the sum of the conduction-band, $d_{{\rm c}}$,
and the valence-band, $d_{{\rm v}}$, dilatation deformation potentials.
For bP, using the elastic constants measured in Ref.~\onlinecite{Yoshizawa_1986}
and discussed in Ref.~\onlinecite{Appalakondaiah_2012}, we have obtained
$d_{{\rm c}}+d_{{\rm v}}\approx$ 4.3~eV. This is about a half the
value calculated in Ref.~\onlinecite{Guan_2016} and the values of
8.19-to-9.89~eV for $d_{{\rm c}}+d_{{\rm v}}$ reported in Ref.~\onlinecite{Yoshizawa_1986}.
Such a discrepancy could be due to intrinsic limitations of the empirical
pseudopotentials we have employed, to uncertainties in the values
of the elastic constants, and to additional `internal' atomic displacements
that occur under strain. On the contrary, for monolayer phosphorene,
we have obtained larger dilatation deformation potentials, $d_{{\rm c}}+d_{{\rm v}}\approx$
6.6~eV. 
Of course, we cannot estimate separately the contribution of the conduction
band and of the valence band (given as a 33\%-66\% split in Ref.~\onlinecite{Asahina_1982}).
However, despite their uncertainty, these results constitute a highly
suggestive argument to explain the larger carrier mobility in thicker
films.
From Figs.~\ref{fig:P_mono_DK} and ~\ref{fig:P_bi_DK}, we observe 
that the effective deformation potential (Eq.~(\ref{eq:Defpot})) for the acoustic modes increases going from mono- to bi-layer. However, as
mentioned earlier, in the case of bilayers, the low-energy optical modes have multiple branch crossings with the acoustic modes at small q-vectors, making it difficult or impossible to
separate the contribution of the acoustic modes. We may argue that the effective deformation potential in acoustic modes might be
much lower than what we see in Fig.~\ref{fig:P_bi_DK} and low-energy optical modes might have a higher
contribution to the scattering rates.

\textit{4. Stiffening of the inter-layer optical modes.} Whereas when moving
from monolayers to bilayers the carrier mobility is kept low by the
presence of low-energy inter-layer optical modes, when moving to thicker
multi-layer systems this effect is expected to be less significant
and even show a qualitatively opposite behavior. Indeed, the amplitude
of these optical modes decreases as their energy increases. This is
the result of their increased stiffness due to the coupling with various
layers, as the energy of the lowest-frequency interlayer mode grows
from $\approx$ 1.5~meV in bilayers to about 25~meV for TO phonons
in bulk black phosphorus\cite{Morita_1986}. The situation is complicated
by the splitting of several modes into `inner' and `surface' modes\cite{Xin-Hu_2016}
in layered structures, but this trend constitutes another possible
cause for a mobility increasing in thicker films.

It would be extremely interesting to study in detail the thickness dependence
of all of these effects. Unfortunately, it is impossible to treat
correctly the 2D-to-3D transition \textendash{} and so, also the $Z$-to-$\Gamma$
direct-band-gap transition and the mobility change \textendash{} when
calculating the band structure and the vibrational properties of many-layer
systems. Indeed, this would require accounting for inelastic, phase-breaking
phonon scattering within a DFT (or even GW) framework, a task obviously still
elusive. Yet, the results discussed here give a qualitative idea of
why the carrier mobility decreases so sharply when moving from bulk
black phosphorus to monolayer phosphorene. In particular, the observation
that {\em phosphorene behaves more like a conventional semiconductor
than a van der Waals material \textendash{} as discussed in item 2
above \textendash{} seems to explain the strong thickness dependence
of the carrier mobility shown in Fig.~4(c) of Ref.~\onlinecite{Liu_2014}
and suggested by Table~\ref{tab:mu_exp}}.\\

\section{Conclusions}

\label{Conclusions} 

The widely scattered theoretical predictions about the carrier mobility
in 2D crystals that have been reported in the literature have prompted
us to analyze critically the reasons for this confusion. Taking monolayer
and bilayer phosphorene as examples of widely studied materials, we
have identified the assumed simplifying isotropy of the electron-phonon
matrix elements, the use of the `band' deformation potential instead
of the proper carrier-phonon matrix elements, and the associated neglect
of the wavefunction-overlap effects as the main sources of this confusion.
Using a simple \textendash{} but, hopefully, not oversimplified \textendash{}
model, we have shown that, unfortunately, the most accurate models
predict the less exciting values for the carrier mobility. These do
not exceed $\approx25$ cm$^{2}$ V$^{-1}$ s$^{-1}$ at 300~K for both
electrons and holes.
We have also employed Monte Carlo simulations \textendash{} based
on a band structure and carrier-phonon scattering rates calculated
using two separate \textit{ab initio} DFT \textendash{} methods to obtain both better estimates
of the low-field mobility in phosphorene monolayers and bilayers,
and information about high-field transport properties. We found that
calculating the carrier-phonon interaction using small dispacements and DFPT yield
similar results. Our study further predicts a decrease in mobility moving from mono- to bi-layers. 
Most important, we have argued (unfortunately only at a qualitative level) that phosphorene, because 
of its lone-pair of `out-of-plane' $p$-electrons,
behaves more like a `conventional' semiconductor than a van der Waals
material, an observation that may explain the thickness dependence
of the carrier mobility reported in the literature.
\acknowledgments 
This work has been supported by Taiwan Semiconductor Manufacturing
Company, Ltd.. Edward Chen acknowledges Jack Sun for management support.



\begin{thebibliography}{100}
\bibitem{Wang_2014}           W.~Zhang, Z.~Huang, W.~Zhang, and Y.~Li,
                              Nano Res. \textbf{7}, 1731 (2014). 
\bibitem{Houssa_2016b}       M.~Houssa, A.~Dimoulas, and A.~Molle eds. (CRC Press -Taylor\&Francis,
                              Boca Raton, Florida, 2016),
                               \textit{2D Materials for Nanoelectronics}.
\bibitem{Molle_2017}          A.~Molle, J.~Goldberger, M.~Houssa, Y.~Xu, S.-C.~Zhang, and D.~Akinwande, 
                              Nature Mat. \textbf{16}, 163 (2017).

\bibitem{DFT_review}          See for example the recent reviews of the history and state of the art of density functional theory by R.~O.~Jones,
                              Rev. Mod. Phys. \textbf{87}, 897 (2015) or by P.~J.~Hasnip, K.~Refson, M.~I.~J.~Probert, J.~R.~Yates, S.~J.~Clark, and C.~J.~Pickard,
                              Phil. Trans. A: Math. Phys. Eng. Sci. \textbf{372}, 20130270 (2014).

\bibitem{VASP1}               G.~Kresse and J.~Hafner, 
                              Phys. Rev. B \textbf{47}, RC558 (1993). 
\bibitem{VASP2}               G.~Kresse, 
                              Ph.~D. thesis, Technische Universit\"{a}t Wien (1993) 
\bibitem{VASP3}               G.~Kresse and J.~Furthm\"{u}ller, 
                              Comput. Mat. Sci. \textbf{6}, 15 (1996). 
\bibitem{VASP4}               G.~Kresse and J.~Furthm\"{u}ller, 
                              Phys. Rev. B \textbf{54}, 11169 (1996).

\bibitem{Giannozzi_2009}      P.~Giannozzi, S.~Baroni, N.~Bonini, M.~Calandra, R.~Car, C.~Cavazzoni, D.~Ceresoli, G.~L.~Chiarotti, M.~Cococcioni,
                              I.~Dabo, \textit{et al.}, 
                              J. Phys.: Condens. Matter \textbf{21}, 395502 (2009).

\bibitem{Vandenberghe_2015}   W.~G.~Vandenberghe and M.~V.~Fischetti,
                              Appl. Phys. Lett. \textbf{106}, 013505 (2015).
\bibitem{Baroni_2001}         S.~Baroni, S.~de~Gironcoli, A.~Dal~Corso, and P.~Giannozzi, 
                              Rev. Mod. Phys. \textbf{73}, 515 (2001). 
\bibitem{Giustino_2017}       F.~Giustino, 
                              Rev. Mod. Phys. \textbf{89}, 015003 (2017)
                              
\bibitem{Ziman_1958}          J.~M.~Ziman, \textit{Electrons and Phonons}
                              (Oxford University Press, Oxford, UK, 1958).    
                              
                              
\bibitem{Zollner_1991}        S.~Zollner, S.~Gopalan, and M,~Cardona,
                              Phys. Rev. B \textbf{44}, 13446 (1991). 
\bibitem{Zollner_1992}        S.~Zollner, S.~Gopalan, and M,~Cardona,
                              Semicond. Sci. Technol. \textbf{7}, B137 (1992).    
                              
\bibitem{Fischetti_1991}      M.~V.~Fischetti and J.~Higman, 
                              in \textit{Monte Carlo Device Simulation: Full Band and Beyond},
                              K.~Hess ed., Kluwer Academic (Norwell, MA, 1991), pp. 123-160. 
\bibitem{Yoder_1993}          P.~D.~Yoder, V.~D.~Natoli and R.~M.~Martin,
                              J. Appl. Phys. \textbf{73}, 4378 (1993). 
                              
\bibitem{Choi_1999}           H.~J.~Choi and J.~Ihm, 
                              Phys. Rev. B \textbf{59}, 2267 (1999). 

\bibitem{Gunst_2016}          T.~Gunst, T.~Markussen, K.~Stokbro, and M.~Brandbyge,
                              Phys. Rev. B \textbf{93}, 035414 (2016).     
                              
\bibitem{Geim_2007}           A.~K.~Geim and K.~S.~Novoselov, 
                              Nature Mat. \textbf{9}, 183 (2007).

\bibitem{Houssa_2011}         M.~Houssa, E.~Scalise, K.~Sankaran, G.~Pourtois, V.~V.~Afanasev, and A.~Stesmans, 
                              Appl. Phys. Lett. \textbf{98}, 223107 (2011). 
\bibitem{Vogt_2012}           P.~Vogt, P.~De~Padova, C.~Quaresima, J.~Avila, E.~Frantzeskakis, M.~C.~Asensio, A.~Resta, B.~Ealet, and G.~Le~Lay,
                              Phys. Rev. Lett. \textbf{108}, 155501 (2012). 
\bibitem{Roome_2014}          Nathanael~J.~Roome and J.~David~Carey, 
                             ACS Appl. Mater. Interfaces \textbf{6}, 7743 (2014). 

\bibitem{Tao_2015}            L.~Tao, E.~Cinquanta, D.~Chiappe, C.~Grazianetti, M.~Fanciulli, M.~Dubey, A.~Molle, and D.~Akinwande, 
                              Nature Nanotechn. \textbf{10}, 227 (2015). 

\bibitem{Li_2013}             X.~Li, J.~T.~Mullen, Z.~Jin, K.~M.~Borysenko, M.~Buongiorno~Nardelli, and K.-W.~Kim, 
                              Phys. Rev. B \textbf{87}, 115418 (2013). 
\bibitem{Davila_2014}         M.~E.~D\'{a}vila, L.~Xian, S.~Cahangirov, A.~Rubio, and G.~Le~Lay (2014), 
                              New J. Phys. \textbf{16}, 095002 (2014).

\bibitem{Gomez_2014}          A.~Castellanos-Gomez, L.~Vicarelli, E.~Prada, J.~Island, K.~L. Narasimha-Acharya, S.~I.~Blanter, D.~J.~Groenendijk,
                              M.~Buscema, G.~A.~Steele, J.~V.~Alvarez, \textit{et al.}, 
                              2D Mater. \textbf{1}, 02500 (2014). 
\bibitem{Xia_2014}            F.~Xia, H.~Wang, and Y.~Jia, 
                              Nature Comm. \textbf{5}, 4458 (2014). 
\bibitem{Li_2014}             L.~Li, Y.~Yu, G.~J.~Ye, Q.~Ge, X.~Ou, H.~Wu, D.~Feng, X.~H.~Chen, and Y.~Xhang, 
                              Nature Nanotechnol. \textbf{9}, 372 (2014). 
\bibitem{Liu_2014}            H.~Liu, A.~T.~Neal, Z.~Zhu, Z.~Luo, X.~Xu, D.~Tom\'{a}nek, and P.~D.~Ye, 
                              ACS Nano \textbf{8}, 4033 (2014). 
\bibitem{Cao_2015}            Y.~Cao, A.~Mishchenko, G.~L.~Yu, K.~Khestanova, A.~Rooney, E.~Prestat, A.~V.~Kretinin, 
                              P.~Blake, M.~B.~Shalom,G.~Balakrishnan, \textit{et al.}, 
                              Nano Lett. \textbf{15}, 4914 (2015). 
\bibitem{Doganov_2015}        R.~A.~Doganov, S.~P.~Koenig, Y.~Yeo, K.~Watanabe, T.~Taniguchi, and B.~\"{O}yilmaz, 
                              Appl. Phys. Lett. \textbf{106}, 083505 (2015). 
\bibitem{Xiang_2015}          .~Xiang, C.~Han, J.~Wu, S.~Zhong, Y.~Liu, J.~Lin, X.-A.~Zhang, W.~P.~Hu, B.~\"{O}zyilmaz, A.~H.~Castro~Neto, A.~Thye, S.~Wee, and W.~Chen, 
                              Nature Comm. \textbf{6}, 6485 (2015). 
\bibitem{Gillgren_2015}       N.~Gillgren, D.~Wickramaratne, Y.~Shi, T.~Espiritu, J.~Yang, J.~Hu, J.~Wei, X.~Liu, Z.~Mao, K.~Watanameb, \textit{et al.}, 
                              2D Mater. \textbf{2}, 011001 (2015). 
\bibitem{Tayari_2015}         V.~Tayari, N.~Hemsworth, I.~Fakih, A.~Favron, E.~Gaufr\'{e}s, G.~Gervais, R.~Martel, and T.~Szkopek, 
                              Nature Comm. \textbf{6}, 7702 (2015).

\bibitem{Kamal_2015}          C.~Kamal and Motohiko~Ezawa, 
                              Phys. Rev. B \textbf{91}, 085423 (2015). 
\bibitem{Li_2016}             Z.~Li, W.~Xu, Y.~Yu, H.~Du, K.~Zhen, J.~Wang, L.~Luo, H.~Qiua, and X.~Yang, 
                              J. Mater. Chem. C \textbf{4}, 362 (2016). 
\bibitem{Pizzi_2016}          G.~Pizzi, M.~Gibertini, E.~Dib, N.~Marzari, G.~Iannaccone, and G.~Fiori, 
                              Nature Comm. \textbf{7}, 12585 (2016). 
\bibitem{Xu_2017}             Y.~Xu, B.~Peng, H.~Zhang, H.~Shao, R.~Zhang, and H.~Zhu 
                                 Annalen der Physik \textbf{529}, 1600152 (2017).             

\bibitem{Singh_2016}          D.~Singh, S.~K.~Gupta, Y.~Sonvane, and I.~Luka\v{c}evi\'{c}
                              J. Mater. Chem. C \textbf{4}, 6386 (2016). 
\bibitem{Ji_2016}             J.~Ji, X.~Song, J.~Liu, Z.~Yan, C.~Huo, S.~Zhang, M.~Su, L.~Liao, W.~Wang, Z.~Ni, Y.~Hao, and H.~Zeng, 
                              Nature Comm. \textbf{7}, 13352 (2016).

\bibitem{Xu_2013}             Y.~Xu, B.~Yan, H.-J.~Zhang, J.~Wang, G.~Xu, P.~Tang, W.~Duan, and S.-C.~Zhang, 
                              Phys. Rev. Lett. \textbf{111}, 136804 (2013). 
\bibitem{Zhu_2015}            F.-f.~ Zhu, W.-J.~Chen, Y.~Xu, C.-L.~Gao, D.-D.~Guan, C.-H.~Liu, D.~Qian, S.~C.~Zhang and J.-F.~Jia, 
                              Nature Mat. \textbf{14}, 1020 (2015). 
\bibitem{Suarez_2015}         A.~Suarez~Negreira, W.~G.~Vandenberghe, and M.~V.~Fischetti, 
                              Phys. Rev. B \textbf{91}, 245103 (2015).
                              
\bibitem{Ford_1992}           W.~K.~Ford, T.~Guo, K.-J.~Wan, and C.~B.~Duke,
                              Phys. Rev. B \textbf{45}, 11896 (1992). 
\bibitem{Whittle_1995}        R.~Whittle, A.~Murphy, E.~Dudzik, I.~T.~McGovern, A,~Hempelmann, C.~Nowak, D.~R.~Zah, A.~Cafolla, and W.~Braun,
                              J. Synchrotron Radiat. \textbf{2}, 256 (1995). 
\bibitem{Khomitsky_2014}      D.~V.~Khomitsky and A.~A.~Chubanov,
                              J. Exp. Th. Phys. \textbf{118}, 457 (2014). 
\bibitem{Reis_2016}           F.~Reis, G.~Li, L.~Dudy, M.~Bauernfeind, S.~Glass, W.~Hanke, R.~Thomale, J.~Sch\"{a}fer, and R.~Claessen,
                              Science \textbf{357}, 287 (2017).
                              
\bibitem{Akahama_1983}        Y.~Akahama, A.~Endo, and S.~Narita, 
                              J. Phys. Soc. Jpn. \textbf{6}, 2148 (1983). 
\bibitem{Morita_1986}         A.~Morita, 
                              Appl. Phys. A \textbf{39}, 227(1986).   
                              
\bibitem{Qiao_2014}           J.~Qiao, X.~Kong, F.~Yang, and W.~Ji, 
                              Nature Comm. \textbf{5}, 4475 (2014).     
                              
\bibitem{Jin_2016}            Z.~Jin, J.~T.~Mullen, and K.~W.~Kim, 
                              Appl. Phys. Lett. \textbf{109}, 053108 (2016).

\bibitem{Trushkov_2017}       Y.~Trushkov and V.~Perebeinos, 
                              Phys. Rev. B \textbf{95}, 075436 (2017).     
                              
\bibitem{Rudenko_2016}        A.~N.~Rudenko, S.~Brener, and M.~I.~Katsnelson,
                              Phys. Rev. Lett. \textbf{116}, 246401 (2016). 

\bibitem{Liao_2015}           B.~Liao, J.~Zhou, B.~Qiu, M.~S.~Dresselhaus, and G.~Chen, 
                              Phys. Rev. B \textbf{91}, 235419 (2015).  
\bibitem{AlTaleb_2016}        A.~Al~Taleb and D.~Farias, 
                              J. Phys.: Condens. Matter \textbf{28}, 103005 (2016). 
\bibitem{Balandin_2014}       A.~Balandin, 
                              MRS Bull. \textbf{39}, 817 (2014).

\bibitem{Ong_2012}            Z.-Y.~Ong and M.~V.~Fischetti, 
                              Phys. Rev. B \textbf{86}, 165422 (2012); Erratum: \textit{ibid} \textbf{86}, 199904(E) (2012).    

\bibitem{Lew_2015}            L.~C.~Lew Yan Voon, J.~Wang, Y.~Zhang, and M.~Willatzen, 
                              J. Phys.: Conf. Series \textbf{663}, 012042 (2015). 
\bibitem{Fukuoka_2015}        Shuhei~Fukuoka, Toshihiro~Taen, and Toshihito~Osada,
                              J. Phys. Soc. Jpn. \textbf{84}, 121004 (2015).
                              
\bibitem{Piscanec_2004}       S.~Piscanec, M.~Lazzeri, F.~Mauri, A.~C.~Ferrari, and J.~Robertson, 
                              Phys. Rev. Lett. \textbf{93}, 185503 (2004). 
\bibitem{Lazzeri_2006}        M.~Lazzeri and F.~Mauri, 
                              Phys. Rev. Lett. \textbf{97}, 266407 (2007). 
\bibitem{Lazzeri_2008}        M.~Lazzeri, C.~Attaccalite, L.~Wirtz, and F.~Mauri, 
                              Phys. Rev. B \textbf{78}, 081406(R) (2008).

\bibitem{Borysenko_2010}      K.~M.~Borysenko, J.~T.~Mullen, E.~A.~Barry, S.~Paul, Y.~G.~Semenov, J.~M.~Zavada, M.~Buongiorno~Nardelli, and K.~W.~Kim,
                              Phys. Rev. B \textbf{81}, 121412(R) (2010).       
                              
\bibitem{Fischetti_2013}      M.~V.~Fischetti, J.~Kim, S.~Narayanan, Z.-Y.~Ong, C.~Sachs, D.~K.~Ferry, and S.~J.~Aboud, 
                              J. Phys.: Cond. Matter \textbf{25}, 473202 (2013).
\bibitem{Nakamura_2017}       Y.~Nakamura, T.~Zhao, J.~Xi, W.~Shi, D.~Wang, and Z.~Shuai,
                              Adv. Electron. Mat., {\bf 3}, 1700143 (2017).  
\bibitem{Takagi_1994}         S.-I.~Takagi, A.~Toriumi, M.~Iwase, and H.~Tango, 
                              IEEE Trans. Electr. Dev. \textbf{41}, 2357 (1994).                                                      
                                                                                                                                                                                                
 \bibitem{Shao_2013}           Z.-G.~Shao, X.-S.~Ye, L.~Yang, and C.-L.~Wang,
                              J. Appl. Phys. \textbf{114}, 093712 (2013). 

\bibitem{Ye_2014}             X.-S.~Ye, X.-G.~Shao, H.~Zhao, L.~Yang, and C.-L.~Wang, 
                             RCS Adv. \textbf{4}, 21216 (2014).   
                             
\bibitem{Fischetti_2016}      M.~V.~Fischetti and W.~G.~Vandenberghe,
                              Phys. Rev. B \textbf{93}, 155413 (2016).   
                              
\bibitem{Bardeen_1950}        J.~Bardeen and W.~Shockley, 
                              Phys. Rev. \textbf{80}, 72 (1950).

\bibitem{Herring_1957}        C.~Herring and E.~Vogt, 
                              Phys. Rev. \textbf{101}, 944 (1957); Erratum: Phys. Rev. \textbf{105}, 1933 (1957). 
\bibitem{Fischetti_1996}      M.~V.~Fischetti and S.~E.~Laux, 
                              J. Appl. Phys. \textbf{80}, 2234 (1996). 
\bibitem{Fischetti_1993}      M.~V.~Fischetti and S.~E.~Laux, 
                              Phys. Rev. B \textbf{48}, 2244 (1993).
                              
\bibitem{note_3}              The fact that this expression does not depend on the density or Fermi energy is the result of the constant density
                              of states in two dimensions for a parabolic dispersion. This implies an energy independent relaxation time. In turn, the energy integral
                              appearing in Eq.~(\ref{eq:mu_gen}), that depends on carrier density (or Fermi energy) via the Fermi-Dirac function 
                              $f(x)=[1+\exp(x-x_{{\rm F}})]^{-1}$ (where $x$ and $x_{{\rm F}}$ are the kinetic energy and Fermi energy in thermal units), becomes
                              proportional to the carrier density $n$ itself, so that 
                              $-\int_{0}^{\infty}{\rm d}x\ x\ \frac{{\rm d}f(x)}{{\rm d}x}\ \left[\int_{0}^{\infty}{\rm d}x\ f(x)\right]^{-1}\ =\ 1$ 
                              becomes independent of the Fermi energy.    

\bibitem{LDA}        J.~P.~Perdew and Y.~Wang,
                              Phys. Rev. B \textbf{45}, 13244 (1992).  
                              
\bibitem{Perdew_1996}         J.~P.~Perdew, K.~Burke, and M.~Ernzerhof,
                              Phys. Rev. Lett. \textbf{77}, 3865 (1977); Erratum Phys. Rev. Lett. \textbf{78}, 1396 (1997).    
                              
\bibitem{Blochl_1994}         P.~E.~Blochl, 
                              Phys. Rev. \textbf{50}, 17953 (1994).                                                                                                                

\bibitem{Grimme_2004}         S.~Grimme, 
                              J. Comp. Chem. \textbf{25}, 46 (2004). 
\bibitem{optPBE}              J.~Klim\v{e}s, D.~R.~Bowler, and A.~Michaelides,
                              J. Phys. Con-dens. Matter \textbf{22}, 022201 (2010). 
                              
                              
\bibitem{optB88}              A.~D.~Becke, 
                              Phys. Rev. A \textbf{88}, 3098 (1988). 
\bibitem{vdW-DF}              M.~Dion, H.~Rydberg, E.~Schr\"{o}der, D.~C.~Langreth, and B.~I.~Lundqvist, 
                              Phys. Rev. Lett. \textbf{92}, 246401 (2004).

\bibitem{PHONOPY}             A.~Togo and I.~Tanaka, 
                              Scr. Mater. \textbf{108}, 1 (2015).                                                                                                                                         
                                                                                 
\bibitem{epw_Giustino}  J.~Noffsinger, F.~Giustino, B.~D.~Malone, C.~H.~Park, S.~G.~Louie and M.~L.~Cohen,
                              Comput. Phys. Commun. \textbf{181}, 2140 (2010).                            
\bibitem{epw_Giustino2} F.~Giustino, M.~ L.~Cohen, and S.~G.~Louie,
                                   Phys. Rev. B \textbf{76}, 165108 (2007).


\bibitem{Gilat_1966}          G.~Gilat and L.~J.~Raubenheimer, 
                              Phys. Rev. \textbf{144}, 390 (1966); Erratum Phys. Rev. \textbf{147}, 670 (1966). 
\bibitem{Fischetti_2011}      M.~V.~Fischetti, B.~Fu, S.~Narayanan, and J.~Kim, 
                              in \textit{Nano-Electronics Devices: Semiclassical and Quantum Transport Modeling}, 
                              D.~Vasileska and S.~M.~Goodnick Eds. (Springer, New York, 2011), pp. 183-247.    
                              
\bibitem{Jacoboni_1983}       C.~Jacoboni and L.~Reggiani, 
                              Rev. Mod. Phys. \textbf{55}, 645 (1983).  
                              
\bibitem{Tran_2014}           V.~Tran, R.~Soklaski, Y.~Liang, and L.~Yang,
                              Phys. Rev. B \textbf{89}, 235319 (2014). 

\bibitem{Zhang_2014}          S.~Zhang, J.~Yang, R.~Xu, F.~Wang, W.~Li, M.~Ghufran, Y.~W.~Zhang, Z.~Yu, G.~Zhang, Q.~Qin, and Y.~Lu,
                              ACS Nano \textbf{8}, 9590 (2014).

\bibitem{Fei_2014a}           R.~Fei, A.~Faghaninia, R.~Soklaski, J.-A.~Yan, C.~Lo, and L.~Yang, 
                              Nano Lett. \textbf{14}, 6393 (2014).
                              

\bibitem{Xin-Hu_2016}         Z.-X.~Hu, X.~Kong, J.~Qiao, B.~Normanda, and W.~ Ji, 
                              Nanoscale \textbf{8}, 2740 (2016).                              
                              
\bibitem{Yan_2008}            J.-A.~Yan, W.~Y.~Ruan, and M.~Y.~Chou, 
                              Phys. Rev. B \textbf{77}, 125401 (2008). 

\bibitem{Park_2008}           C.-H.~Park, F.~Giustino, M.~L.~Cohen, and S.~G.~Louie, 
                              Nano Lett. \textbf{8}, 4229 (2008).

\bibitem{Bellaiche_1996}      L.~Bellaiche, S.-H.~Wei, and A.~Zunger,
                              Phys. Rev. B \textbf{54}, 17568 (1968).
                              
\bibitem{note_1}              The same pseudopotential yields a bandgap of $\approx$ 0.6~eV for bulk black phosphorus when using the same monolayer in-plane
                              lattice constants, of $\approx$ 0.2~eV when using the bulk lattice constants given by H.~Asahina, K.~Shindo, and A.~Morita, 
                              J. Phys. Soc. Jpn. \textbf{51}, 1193 (1982): $a$ = 3.3134~\AA, $b$ = 10.478~\AA, and $c$ = 4.3763~\AA. 
                              These values for the bandgap are respectively larger and smaller than the experimental value ($\approx0.3$ eV from 
                              Ref.~\onlinecite{Akahama_1983}), presumably a result of the strong sensitivity of the calculated bandgap
                              on small variations of the lattice constants. The values for the effective mass reported in Tables~\ref{tab:emass} and 
                              \ref{tab:hmass} for bulk bP have been obtained using the bulk lattice constants.

\bibitem{Rodin_2014}          A.~S.~Rodin, A.~Carvalho, and A.~H.~Castro~Neto,
                              Phys. Rev. Lett. \textbf{112} 176801 {[}2014). 
\bibitem{Li_2014a}            P.~Li and I.~Appelbaum, 
                              Phys. Rev. B \textbf{90} 115439 (2014). 
\bibitem{Peng_2014}           X.~Peng, Q.~Wei, and A.~Copple, 
                              Phys. Rev. B \textbf{90} 085402 (2014).                              

\bibitem{Fischetti_2013a}     M.~V.~Fischetti, B.~Fu, and W.~G.~Vandenberghe,
                              IEEE Trans. Electron Devices \textbf{60}, 3862 (2013).

\bibitem{Charlier_1994}       J.-C.~Charlier, X.~Gonze and J.-P.~Michenaud,
                              Europhys. Lett. \textbf{28}, 403 (1994).

\bibitem{Bjorkman_2012}       T.~Bj\"{o}rkman, A.~Gulans, A.~V.~Krasheninnikov, and R.~M.~Nieminen, 
                              Phys. Rev. Lett. \textbf{108}, 235502 (2012).   
                              
\bibitem{Pierson_1993}        H.~O.~Pierson, \textit{Handbook of Carbon, Graphite, Diamond and Fullerenes - Properties, Processing and Applications},
                              (Noyes Publications, Park Ridge, New Jersey, 1993). 
\bibitem{Dresselhaus_2002}    M.~S.~Dresselhaus and G.~Dresselhaus,
                              Adv. Phys. \textbf{51}, 1 (2002).   
                              
\bibitem{Price_1981}          P.~J.~Price, 
                              Ann. Phys. \textbf{133}, 217 (1981).

\bibitem{Fei_2014}            R.~Fei and L.~Yang, 
                              Nano Lett. \textbf{14}, 2884 (2014). 
\bibitem{Yoshizawa_1986}      M.~Yoshizawa, S.~Endo, Y.~Akahama, and S.~Narita, 
                              J. Phys. Soc. Jpn. \textbf{55}, 1196 (1986). 
\bibitem{Appalakondaiah_2012} S.~Appalakondaiah, G.~Vaitheeswaran, S.~Leb\`{e}gue, N.~Christensen, and A.~Svane, 
                             {\it Effect of van der Waals interactions on the strucfural and elastic proprties of black phosphorus},
                              Phys. Rev. B \textbf{86}, 035105 (2012). 
\bibitem{Guan_2016}           J.~Guan, W.~Song, L.~Yang,  D.~Tom\'{a}nek,
                             {\it Strain-controlled fundamental gap and structure of bulk black phosphorus},
                              Phys. Rev. B \textbf{94}, 045414 (2016). 
                              
\bibitem{Asahina_1982}        H.~Asahina, K.~Shindo, and A.~Morita, J. Phys. Soc. Jpn. \textbf{51}, 1193 (1982).

                                                                                                                                                                           
\end{thebibliography}
\end{document}